\PassOptionsToPackage{table}{xcolor}
\documentclass{article} %
\usepackage[final]{colm2026_conference}
\usepackage{amsmath}
\usepackage{graphicx}
\usepackage{microtype}
\usepackage[hyphens]{url}
\usepackage{hyperref}
\usepackage{booktabs}
\usepackage{xspace}
\usepackage{multirow}
\usepackage{array}
\usepackage{booktabs}
\usepackage{enumitem}

\newcommand{\up}[1]{\raisebox{-0.3ex}{\scriptsize\textcolor{red}{\(\uparrow\)#1}}}
\newcommand{\down}[1]{\raisebox{-0.3ex}{\scriptsize\textcolor{green!50!black}{\(\downarrow\)#1}}}

\usepackage{pgf} %
\usepackage{lineno}
\usepackage[most]{tcolorbox}
\definecolor{darkblue}{rgb}{0, 0, 0.5}
\hypersetup{colorlinks=true, citecolor=darkblue, linkcolor=darkblue, urlcolor=darkblue}
\definecolor{my_lightblue}{RGB}{194, 213, 247}

\title{The Devil Is in the Interface: Evaluating How Tool Architecture Shapes Coding Agent Behavior}

\author{Xiangzhe Xu$^{1*}$
\And
Hamidreza Saghir$^2$
\And
Qianhui Wu$^2$
\And
Marc-Alexandre Côté$^2$
\AND
Tong Wang$^2$
\And
Kiran Lakkaraju$^2$
\And
Kexin Pei$^3$
\And
Xiangyu Zhang$^1$ \AND
{\normalfont $^1$Purdue University \quad $^2$Microsoft Research \quad $^3$The University of Chicago}\\
\{xzx, xyzhang\}@purdue.edu\quad kpei@uchicago.edu \\ 
\{hsaghir, qianhuiwu, macote, tong.wang, kiran.lakkaraju\}@microsoft.com
\\
$^*$ Contribution during an internship at Microsoft Research.
}

\begin{document}

\ifcolmsubmission
\linenumbers
\fi

\maketitle

\begin{abstract}
As large language models continue to improve, agentic systems are becoming increasingly important, and tools are a key design dimension because they determine how agents access information and take action in their environments. 
Prior work on agent tooling has primarily focused on expanding what
agents can do, but has paid less systematic attention to how those
capabilities are organized and exposed to the model. We refer to this
latter design dimension as \emph{tool architecture}.
We study tool architecture in coding agents through controlled
experiments on repository-level issue fixing, comparing six tool
architectures that hold the underlying information and actions similar
while varying how they are organized and exposed to the model, across
three actors and a total of 11{,}700 trajectories.
Our experiments show that, even when tools provide similar
capabilities, tool architecture changes agent behavior:
Compared to a basic architecture where the agent has only the bash
tool,
more structured low-level interfaces improve consistency across
repeated attempts by up to 4.7$\times$; natural-language search broadens repository
exploration and increases access to relevant files by more
than 11\%; and Python CodeAct-style
interfaces achieve similar task performance with 41.6\% fewer steps and
56.3\% lower token usage. By contrast, lightweight text-based cognitive-scaffolding tools, such as tools that let the
agent record intermediate reasoning, have limited effect on actor
behavior.

\end{abstract}

\section{Introduction}

As large language models continue to improve, agentic systems are becoming increasingly important~\citep{work2025,gap2025,state2025,enpower2025}. Beyond continued model scaling, growing attention has turned to harness engineering~\citep{ant2026demystify,oai2026harness,ant2026harness}, which studies how the surrounding system enables a model to interact with external environments and complete complex tasks. Within this broader space, tool design is a particularly important dimension~\citep{NEURIPS2024_8a75ee6d,yang2024swe,yao2024taubenchbenchmarktoolagentuserinteraction,wang2025mcpbenchbenchmarkingtoolusingllm} because tools define how an agent accesses information and interacts with the environment.~\looseness=-1

Most prior work on agent tools has focused on expanding what the agent
can do by giving it new capabilities. In coding agents, this often
means adding components such as program analysis modules~\citep{zhang2024autocoderover,ruan2025specrover}, mechanisms for
rolling out and comparing multiple candidate solutions~\citep{ma2024understand,gao2025trae}, or precomputed
semantic embeddings that help the agent retrieve relevant code~\citep{ma2024understand,ouyang2024repograph,liu2025reposcope}. These
additions improve performance by giving the agent access to information
or actions that would otherwise be unavailable or difficult to obtain.

However, tool capability is not the whole story~\citep{gloaguen2026evaluating,garg2026debug2fix}. Even when two agents can
access similar information and take similar actions, the way those
capabilities are organized and exposed to the model can still
meaningfully affect behavior~\citep{yang2024swe}. It can influence how consistently the
agent solves the same task across runs~\citep{rabanser2026towards}, how broadly or narrowly it
explores a repository~\citep{gloaguen2026evaluating,li2026contextbench}, and how efficiently it uses steps and tokens~\citep{lulla2026impact}. We
refer to this design dimension as \emph{tool architecture}.

\begin{figure}[t]
    \centering
    \includegraphics[width=0.9\linewidth]{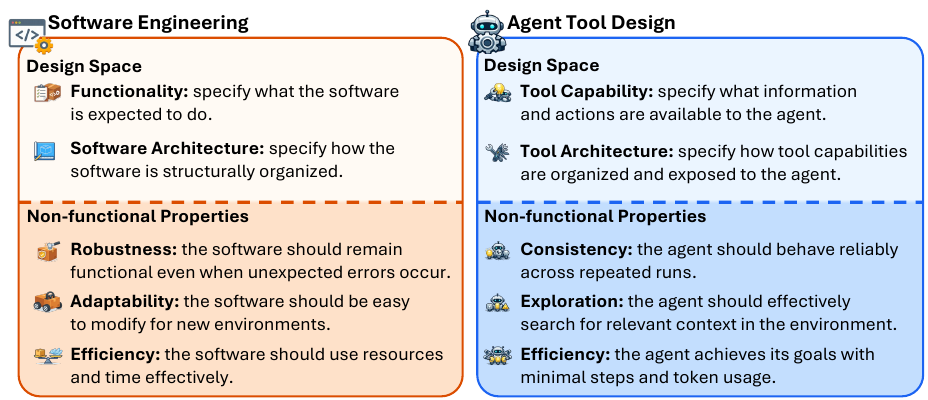}
    \caption{Analogy between software engineering~(SE) and agent tool design. In SE, software architecture is complementary to functionality and
shapes non-functional properties such as robustness~\citep{randell1975system,parnas1972criteria,williams1998performance}. Our study focuses
on the analogous question in agent tool design. We find that, for tools
with similar capabilities, different tool architectures, that is,
different ways of organizing and exposing information and actions to the
model, can affect consistency, exploration, and
efficiency.}
    \label{fig:se-design}
    \vspace{-5pt}
\end{figure}

Figure~\ref{fig:se-design} illustrates our conceptual analogy to
software engineering~\citep{parnas1972criteria}. Tool capability is analogous to software
functionality: it describes what information and actions are available
to the agent. Tool architecture, by contrast, is analogous to software
architecture or interface design~\citep{perry1992foundations,kazman1994toward}: it describes how those capabilities
are organized and exposed to the model. Our experiments show that this
distinction matters in practice for agent tools. Even when two tool
setups support similar tasks, they can still lead the model to behave
differently because they present similar underlying capabilities through
different interfaces.

In practice, however, capability and architecture are often entangled~\citep{garg2026debug2fix,ouyang2024repograph}.
For example, a tool for semantic retrieval over a code repository both
introduces additional capability through embedding-based retrieval and
changes the search architecture by replacing lower-level grep-style
interaction with a more abstract natural-language interface. As a
result, if such a tool improves performance, it is unclear where the
gain should be attributed. Our experiments suggest that part of the
gain can arise from architecture alone: even without embeddings,
implementing semantic search through bash operations but exposing it
through a more natural-language-oriented retrieval interface leads
agents to explore more than 11\% more files in the repository.

To isolate the effect of tool architecture, we design six tool
architectures based on commonly adopted tool designs in existing coding agents~\citep{openhands,claudecode,gao2025trae,antoniades2024swe,smolagents,yang2024swe,augcode} and implement them so that
they expose similar information and similar capability to interact with
the environment. We study two dimensions of architecture. The first is
the \emph{level of abstraction}: some interfaces expose low-level
operations~(e.g., file editing) directly, some provide natural-language search interfaces,
and some let the agent write executable code instead of issuing
individual tool calls. 
The second is \emph{cognitive scaffolding}: those tools provide interfaces
 for recording intermediate reasoning or
task-specific assumptions. Here are our key contributions
and findings:
\begin{itemize}[itemsep=1pt]
    \item We design controlled experiments for tool architecture by
    implementing six representative architectures with similar
    capabilities.
    \item We evaluate three actor models over 11{,}700 coding-agent
    trajectories in total. The experiments find that when tool capabilities are kept similar, overall task performance remains broadly similar across architectures.
    \item More structured low-level interfaces improve
    repeated-attempt consistency, by up to 4.7$\times$. Natural-language search interfaces broaden repository exploration and increase access to plausibly relevant files by more
    than 11\%, but also introduce additional noise. Code-execution-based interfaces achieve similar task
    performance with 41.6\% fewer steps and 56.3\% lower token cost.~\looseness=-1
    \item By contrast, lightweight text-based cognitive-scaffolding tools have
    limited effect because they rarely change the actors' underlying
    reasoning behavior.
    \item We release our code and data at \url{https://github.com/XZ-X/tool-arch-study.git}.
\end{itemize}

\section{Setups}

\subsection{Tool Setups}

We study six tool architectures that are commonly adopted in existing
coding agents; Table~\ref{tab:tool-taxonomy} summarizes the six setups
and representative agents with corresponding design choices.
Our goal is to study tool architecture, so we design these setups to keep their
capabilities as similar as possible. We use \texttt{BashOnly} as the
baseline, where the agent interacts with the repository only through a
general-purpose shell. From this baseline, we vary tool architecture
along two dimensions. 

The first dimension is the \emph{level of abstraction}: \texttt{Atomic}
exposes common low-level operations as explicit tools, \texttt{NLSearch}
provides a natural-language search interface for repository exploration, and
\texttt{Python} lets the agent write executable Python code instead of issuing
individual tool calls.

The second is \emph{cognitive scaffolding}: \texttt{HypoTrack} lets the
agent record and update hypotheses about the task, while
\texttt{Scratchpad} lets the agent write down intermediate reasoning
during problem solving. Note that we implement these scaffolds as lightweight text-based tools
so that they provide an explicit channel for actor-written intermediate state without adding retrieval~\citep{ouyang2024repograph}, memory management~\citep{wang2026improvingcodelocalizationrepository}, new task information~\citep{ruan2025specrover}, or enforcing a different reasoning policy~\citep{antoniades2024swe}. This design lets us isolate whether simple cognitive-scaffolding interfaces change actor behavior under otherwise similar capabilities.

\begin{figure}[t]
    \centering
    \includegraphics[width=0.95\linewidth]{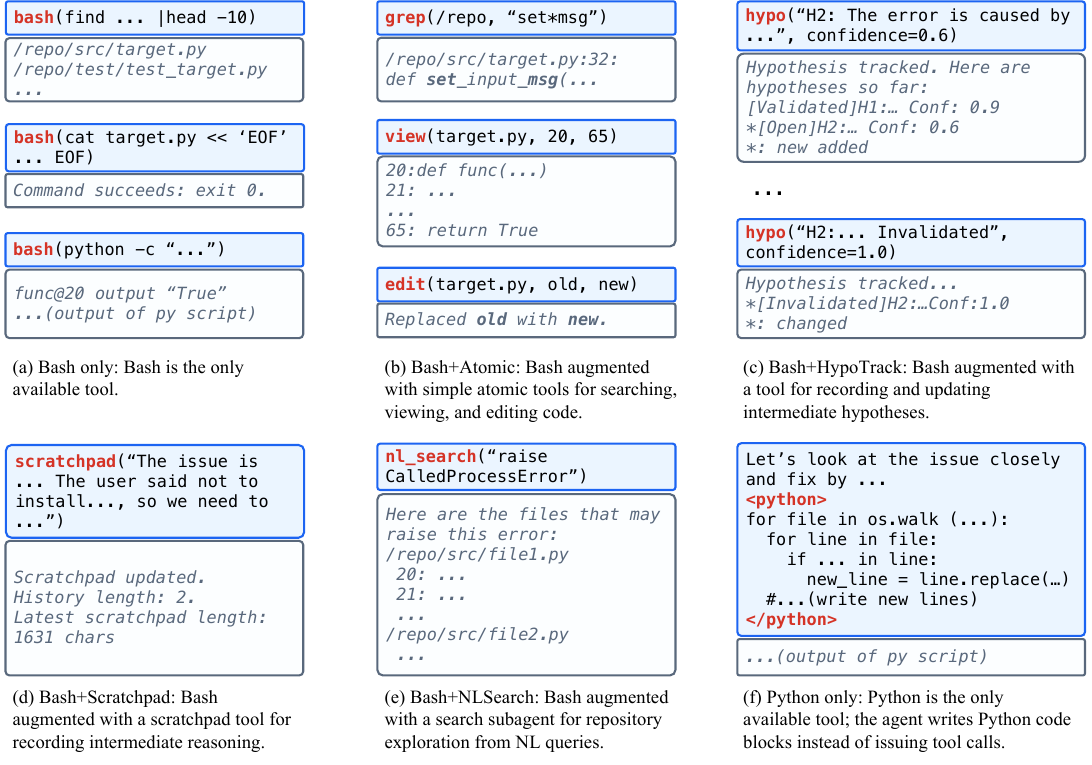}
    \caption{Examples of tool architectures evaluated in this work. The panels illustrate how similar capabilities are exposed to the agent through different interfaces in each setup.}
    \label{fig:tool-setup}
\end{figure}

Figure~\ref{fig:tool-setup} shows concrete examples of the interaction style in each setup.

\noindent \textbf{Baseline: BashOnly.}
The baseline setup exposes only a general-purpose bash interface. The agent interacts with the repository through unconstrained shell commands for navigation, search, file inspection, editing, and execution. This setup offers broad flexibility but little structure: the model must compose low-level commands to carry out even routine subtasks. We use it as the reference point because it provides a minimally structured yet widely used~\citep{xia2025live,jimenez2023swe,yang2024swe} coding-agent environment.

\noindent \textbf{Level of abstraction: Atomic.} The
\textit{Atomic} setup augments bash with a small collection of simple
tools for common localized operations, such as repository search,
bounded file viewing, targeted string replacement, and file creation.
We include this setup because prior work~\citep{yang2024swe} suggests that clear, simple
tool interfaces can improve agent performance, and such structured
high-frequency primitives are common~\citep{openhands,claudecode,yang2024swe,gao2025trae} in mainstream coding agents. The \textit{Atomic} setup does not extend bash capability; it only
repackages common shell actions into simple tools (see
Table~\ref{tab:atomic-tools} in the Appendix for the mapping between atomic tools we adopted and
their bash equivalents).

\noindent \textbf{Level of abstraction: NLSearch.} The
\textit{NLSearch} setup augments bash with a tool that takes a
natural-language query as input and returns relevant code snippets. We
include this setup because it captures the interface of commonly used
semantic search capabilities~\citep{augcodectx,liu2025reposcope,ma2024understand} in coding agents, which let the model express
retrieval needs at a higher level than raw shell commands. To avoid
exceeding the capability of bash, however, our implementation does \textbf{\textit{not}}
use semantic indexing or embedding-based retrieval. Instead, we
implement \textit{NLSearch} as a subagent with the same actor model,
which takes a natural-language query and uses bash commands such as
\texttt{grep} to iteratively search for and return potentially relevant
snippets.

\noindent \textbf{Level of abstraction: Python.} The
\textit{Python} setup replaces explicit tool calls with executable
Python blocks: instead of invoking bash or specialized tools, the agent
writes Python code to traverse the repository, inspect files, edit
content, and run checks. We include this setup because code execution has become a common agent interface, especially in CodeAct-style systems~\citep{wang2024executable,openhands,smolagents}.
This setup also does not exceed the capability of bash in practice, since agents with shell
access can already write and execute Python scripts to carry out the
same operations; the difference is therefore one of interaction style
rather than underlying capability.
Appendix~\ref{appdx:efficiency-details} further validates this design choice:
in a manual inspection of 100 sampled Python actions, 97\% correspond to
operations already available in \texttt{BashOnly}, either as direct bash-side
operations or as Python scripts that \texttt{BashOnly} actors commonly create
and run from bash.

\newcolumntype{L}[1]{>{\raggedright\arraybackslash}p{#1}}
\newcolumntype{M}[1]{>{\centering\arraybackslash}m{#1}}

\begin{table}[t]
\centering
\scriptsize
\caption{A taxonomy of the tool architectures studied in this work and representative
coding agents with corresponding tool designs.}
\setlength{\tabcolsep}{5pt}
\renewcommand{\arraystretch}{1.25}
\begin{tabular}{M{0.20\linewidth} M{0.20\linewidth} L{0.52\linewidth}}
\toprule
\textbf{Category} & \textbf{Tool Arch.} & \textbf{Representative Agents} \\
\midrule
\multirow{3}{*}{Baseline}
& BashOnly
& SWE-Bench bash only leaderboard~\citep{jimenez2023swe}, Live-SWE-agent~\citep{xia2025live}, Mini-SWE-Agent~\citep{yang2024swe}. \\
\midrule
\multirow{4}{*}{Level of Abstraction}
& Atomic
& OpenHands~\citep{openhands}, Claude Code~\citep{claudecode}, TRAE~\citep{gao2025trae}, SWE-Agent~\citep{yang2024swe}. \\
& NLSearch
& Augment Code context engine~\citep{augcodectx}. \\
& Python
& Smolagents~\citep{smolagents}. \\
\midrule
\multirow{5}{*}{Cognitive Scaffolding}
& HypoTrack
& Sequential thinking tool~\citep{gao2025trae,claudecode,openhands,augcode}, SWE-Search~\citep{antoniades2024swe}. \\
& Scratchpad
& Sequential thinking tool~\citep{gao2025trae,claudecode,openhands,augcode}. \\
\bottomrule
\end{tabular}
\label{tab:tool-taxonomy}
\end{table}

\noindent \textbf{Cognitive scaffolding: HypoTrack.} The
\textit{HypoTrack} setup adds a simple hypothesis-tracking tool to the
bash environment that lets the agent explicitly record a hypothesis and
its confidence, and later update the hypothesis status.
We include this tool because mainstream agents use similar
\texttt{thinking} tools to track branches~\citep{gao2025trae,claudecode}, and because coding agents
often benefit from exploring multiple paths~\citep{gao2025trae}, though such behavior is
usually induced by broader system design~\citep{antoniades2024swe} rather than a single tool call. We want to isolate
whether that lightweight scaffolding alone can encourage broader
exploration.
To keep \textit{HypoTrack} aligned with the capability of bash, the
tool only records a hypothesis and its status, both of which could
also be expressed directly by a bash-only agent in its reasoning text.

\noindent \textbf{Cognitive scaffolding: Scratchpad.} The
\textit{Scratchpad} setup adds a simple tool that lets the agent output
free-form thinking content. We include it because similar scratchpad-style
interfaces are common in mainstream agents~\citep{gao2025trae,claudecode,openhands,augcode}. To keep \textit{Scratchpad} aligned with the capability of bash, the
tool only accepts free-form text and returns a confirmation message,
adding no capability beyond what could already be written in the
agent's reasoning text.

\subsection{Evaluation Protocol}
\vspace{-5pt}

We evaluate agents on issue-resolution tasks from SWE-bench Live~\citep{zhang2025swe}, a contamination-controlled benchmark for repository-level bug fixing. Each instance provides an issue description and a code repository snapshot, and the agent must generate a patch that resolves the issue. Outputs are evaluated against hidden human-authored test cases, which are not exposed to the agent.

To make repeated evaluation tractable under our budget, we construct a
subset of SWE-bench Live by randomly sampling 25 of its 100
repositories and retaining at most five issues per repository, yielding
65 problem instances in total. For every actor--setup pair, we execute 10 independent rollouts per instance (i.e., attempt at solving the task). Using repeated rollouts lets us measure not only aggregate task performance, but also behavioral properties such as consistency across reruns and diversity in exploration and final solutions.

We evaluate three actor models: Qwen3Coder-30B~\citep{yang2025qwen3technicalreport}, Kimi K2.5~\citep{kimiteam2026kimik25visualagentic}, and
Claude Sonnet 4.5~\citep{sonnet}. 
We choose them to cover a diverse set of coding agents, including open-weight
and proprietary access, different model scales, and coding- and agent-oriented
training.

We also evaluate whether the main tool-architecture effects extend beyond this
SWE-bench Live subset. Specifically, we consider additional coding tasks covering
issue resolving, feature implementation, and debugging. We use SWE-bench
Verified~\citep{jimenez2023swe} for issue resolving and
SWE-bench Pro~\citep{deng2025swebenchproaiagents} for feature implementation.
For debugging, we construct a separate SWE-bench split by selecting instances
whose issue descriptions contain full stack traces. Due to budget constraints,
we run this additional study with the two open-weight actors,
\texttt{Qwen3Coder-30B} and \texttt{Kimi-K2.5}, and focus on the pairwise
comparisons most directly tied to our main findings. For each
actor--task--setup combination, we run 5 independent rollouts per instance.
Appendix~\ref{appendix:generalization-more-tasks} reports the detailed results.

\newcommand{\consk}[1]{\ensuremath{{\fontencoding{T1}\selectfont\texttt{pass\textasciicircum #1}}\xspace}}

\subsection{Evaluation Dimensions and Metrics}
\vspace{-5pt}
We evaluate tool architectures along four complementary dimensions:
\emph{task resolve rate}, \emph{consistency}, \emph{exploration}, and
\emph{efficiency}. Below, we describe the motivation and intuition of each metric, and discuss the formal definitions in
Appendix~\ref{appdx:metric-definitions}.

\noindent \textbf{Task resolve rate.}
We first report the task resolve rate as the basic measure of whether an
agent solves a task. It captures the fraction of repeated attempts that
successfully resolve the benchmark problem instance, averaged across instances.
This metric tells us how effective a setup is overall.

\noindent \textbf{Consistency.}
A reliable coding agent should not only solve a task, but should achieve
success consistently across repeated attempts. Consistency matters
because users care about whether the same setup can be trusted to
reproduce success, rather than exhibiting success only intermittently~\citep{gupta2026reliabilitybench,rabanser2026towards}.
This is especially important in longer-horizon coding tasks, where even
a small mistake can cascade into later decisions. 
We therefore measure consistency by asking how often success
remains stable across multiple repeated attempts of the same problem instance. The metric is typically denoted as \consk{k}~\citep{yao2024taubenchbenchmarktoolagentuserinteraction}.
In our experiments, we report this metric for repeated groups
of size \(k \in \{5,7,9\}\).

\noindent \textbf{Exploration.}
Recent work suggests that coding agents often struggle to acquire
repository-level context effectively and may under-explore the codebase
on repository-grounded tasks~\citep{li2026contextbench}. Motivated by
this observation, we study whether different tool architectures induce
different exploration behavior. In particular, we ask whether some tool
architectures encourage the agent to explore more diverse parts of the
repository across repeated attempts and to produce different candidate
solutions. We measure exploration from two complementary perspectives: how much
repeated attempts differ in the files they read, and how much their
final patches differ from one another. Concretely, we use Jaccard~\citep{jaccard1901etude}
distance to measure diversity in file access and CodeBLEU~\citep{ren2020codebleu} distance to
measure diversity in generated solutions; see Appendix~\ref{appdx:metric-definitions} for details.

\noindent \textbf{Efficiency.}
Besides consistency and exploration, the cost of a coding agent is also
important in real deployment~\citep{bui2026building,xiao2025improving,pan2025hidden,lulla2026impact}.
We measure efficiency using three trajectory-level statistics: input
tokens, output tokens, and trajectory length~(i.e., the number of interaction steps). 
We report the average of
each quantity for every actor--setup pair and interpret these
measurements jointly with task performance, since lower token usage and
fewer steps are not necessarily preferable if they come at the cost of
substantially worse task resolution.

\section{Evaluation Results}

\vspace{-5pt}
\subsection{Task Resolve Rate}
\vspace{-5pt}

The overall task resolve rate is similar across tool architectures for a
given actor model. This is expected, since we intentionally design the
tool architectures to minimize capability differences relative to the
BashOnly setup, allowing us to isolate the effects of architecture on
non-functional properties. See Appendix~\ref{appdx:perf} for details.

\subsection{Consistency}
\label{sec:eval:consistency}
\vspace{-5pt}
\newcolumntype{C}{>{\centering\arraybackslash}p{2.25cm}}
\renewcommand{\arraystretch}{1.08}
\setlength{\tabcolsep}{5pt}

\newcommand{\basecell}[1]{%
  \cellcolor[rgb]{0.97,0.98,0.99}\strut #1%
}

\newcommand{\actormaxdelta}{0.08}
\newcommand{\setactormax}[1]{\gdef\actormaxdelta{#1}}

\newcommand{\deltacell}[3]{%
  \begingroup
  \pgfmathsetmacro{\s}{min(abs(#1)/\actormaxdelta,1)}%
  \ifdim #1pt > 0pt
    \pgfmathsetmacro{\r}{1 - 0.16*\s}%
    \pgfmathsetmacro{\g}{1 - 0.10*\s}%
    \pgfmathsetmacro{\b}{1 - 0.02*\s}%
  \else\ifdim #1pt < 0pt
    \pgfmathsetmacro{\r}{1 - 0.02*\s}%
    \pgfmathsetmacro{\g}{1 - 0.14*\s}%
    \pgfmathsetmacro{\b}{1 - 0.14*\s}%
  \else
    \pgfmathsetmacro{\r}{1}%
    \pgfmathsetmacro{\g}{1}%
    \pgfmathsetmacro{\b}{1}%
  \fi\fi
  \edef\temp{\noexpand\cellcolor[rgb]{\r,\g,\b}\noexpand\strut\unexpanded{#2\;(#3)}}%
  \temp
  \endgroup
}

\begin{table}[t]
\caption{Consistency results across actor models. Each cell reports absolute \consk{k}; parenthesized values denote the change relative to \texttt{BashOnly}. Background color indicates change relative to \texttt{BashOnly}: blue = improvement, orange = decline, gray = baseline.
}
\label{tab:consistency-all-actors}
\centering
\small
\begin{tabular}{llCCC}
\toprule
Actor & Setup & \consk{5} & \consk{7} & \consk{9} \\
\midrule

\setactormax{0.074}
\multirow{6}{*}{\texttt{Qwen3Coder-30B}}
& BashOnly
& \basecell{0.046}
& \basecell{0.031}
& \basecell{0.020} \\
& Atomic
& \deltacell{0.059}{0.106}{\textbf{+0.059}}
& \deltacell{0.067}{0.097}{\textbf{+0.067}}
& \deltacell{0.074}{0.094}{\textbf{+0.074}} \\
& NLSearch
& \deltacell{0.005}{0.051}{+0.005}
& \deltacell{0.008}{0.039}{+0.008}
& \deltacell{0.012}{0.032}{+0.012} \\
& HypoTrack
& \deltacell{-0.006}{0.040}{-0.006}
& \deltacell{0.002}{0.032}{+0.002}
& \deltacell{0.011}{0.031}{+0.011} \\
& Scratchpad
& \deltacell{-0.017}{0.029}{-0.017}
& \deltacell{-0.010}{0.021}{-0.010}
& \deltacell{-0.003}{0.017}{-0.003} \\
& Python
& \deltacell{-0.030}{0.016}{-0.030}
& \deltacell{-0.025}{0.006}{-0.025}
& \deltacell{-0.018}{0.002}{-0.018} \\
\midrule

\setactormax{0.03}
\multirow{6}{*}{\texttt{Kimi-K2.5}}
& BashOnly
& \basecell{0.290}
& \basecell{0.277}
& \basecell{0.266} \\
& Atomic
& \deltacell{0.014}{0.304}{\textbf{+0.014}}
& \deltacell{0.013}{0.289}{\textbf{+0.013}}
& \deltacell{0.014}{0.280}{\textbf{+0.014}} \\
& NLSearch
& \deltacell{-0.031}{0.258}{-0.031}
& \deltacell{-0.031}{0.246}{-0.031}
& \deltacell{-0.031}{0.235}{-0.031} \\
& HypoTrack
& \deltacell{-0.008}{0.282}{-0.008}
& \deltacell{-0.005}{0.272}{-0.005}
& \deltacell{-0.002}{0.265}{-0.002} \\
& Scratchpad
& \deltacell{-0.022}{0.268}{-0.022}
& \deltacell{-0.020}{0.257}{-0.020}
& \deltacell{-0.017}{0.249}{-0.017} \\
& Python
& \deltacell{-0.024}{0.265}{-0.024}
& \deltacell{-0.021}{0.256}{-0.021}
& \deltacell{-0.017}{0.249}{-0.017} \\
\midrule

\setactormax{0.043}
\multirow{6}{*}{\texttt{Sonnet-4.5}}
& BashOnly
& \basecell{0.296}
& \basecell{0.270}
& \basecell{0.252} \\
& Atomic
& \deltacell{0.017}{0.313}{+0.017}
& \deltacell{0.027}{0.297}{+0.027}
& \deltacell{0.031}{0.283}{+0.031} \\
& NLSearch
& \deltacell{0.018}{0.314}{+0.018}
& \deltacell{0.033}{0.303}{\textbf{+0.033}}
& \deltacell{0.043}{0.295}{\textbf{+0.043}} \\
& HypoTrack
& \deltacell{0.020}{0.315}{\textbf{+0.020}}
& \deltacell{0.021}{0.291}{+0.021}
& \deltacell{0.018}{0.271}{+0.018} \\
& Scratchpad
& \deltacell{-0.017}{0.279}{-0.017}
& \deltacell{-0.010}{0.260}{-0.010}
& \deltacell{-0.003}{0.249}{-0.003} \\
& Python
& \deltacell{-0.013}{0.283}{-0.013}
& \deltacell{-0.004}{0.266}{-0.004}
& \deltacell{0.000}{0.252}{0.000} \\
\bottomrule
\end{tabular}
\end{table}

Table~\ref{tab:consistency-all-actors} shows a clear consistency
result: \texttt{Atomic} is the only setup that improves \consk{k}
relative to \texttt{BashOnly} for all three actors. Because our goal
is to isolate the effect of tool architecture within each actor, the
key comparisons are the changes from \texttt{BashOnly} for the same
actor, rather than the absolute \consk{k} values alone. 
We can see that \texttt{Atomic} improves \consk{5}, \consk{7}, and \consk{9} for
\texttt{Qwen3Coder-30B}, \texttt{Kimi-K2.5}, and
\texttt{Sonnet-4.5}, making it the only setup with a uniformly
positive consistency effect.

The magnitude of this gain is also actor-dependent. The improvement is
largest for the weakest actor, \texttt{Qwen3Coder-30B}, where
\texttt{Atomic} increases \consk{k} by 0.059--0.074 depending on
\(k\). The gains are smaller but still positive for the stronger actors:
0.013--0.014 for \texttt{Kimi-K2.5} and 0.017--0.031 for
\texttt{Sonnet-4.5}. This pattern suggests that packaging recurring
low-level repository actions into clearer, more structured primitives is
especially helpful for weaker actors, whose trajectories appear more
vulnerable to small execution mistakes that can accumulate across an attempt.
For the two open-weight actors, we additionally evaluate
\texttt{BashOnly} and \texttt{Atomic} in Appendix~
\ref{app:more-consistency-repeats} using extended repeated evaluation
with three times more attempts, and the results show that these gains
remain stable.

The remaining setups do not show the same cross-actor regularity.
\texttt{NLSearch} and \texttt{HypoTrack} improve consistency for some
actors but reduce it for others, while \texttt{Scratchpad} and
\texttt{Python} are generally neutral to negative. By comparison,
\texttt{Atomic} is the only setup with uniformly positive
\consk{k} deltas across all three actors, and its gains are largest for
\texttt{Qwen3Coder-30B}.

\noindent
\textbf{Mechanism: Atomic improves consistency by reducing interaction
errors.}
A plausible mechanism for \texttt{Atomic}'s consistency gains is that it
reduces low-level environment-interaction errors. This interpretation is
motivated by prior work in system reliability and software
dependability, which often analyzes failures through chains linking
faults, errors, and observed failures~\citep{avizienis2004basic,gupta2026reliabilitybench,rabanser2026towards}. 
In our experiments, \texttt{Atomic} is
associated with fewer such errors than \texttt{BashOnly}, especially
for the weakest actor, and most of the reduction comes from error types that a more structured interface should directly help avoid, such as malformed commands and
broken edits. This pattern suggests that
packaging recurring low-level actions into explicit primitives can
improve stability across repeated attempts by reducing execution-level
mistakes. Appendix~\ref{appdx:consistency-mechanism} provides the full
error analysis and category breakdown.

\begin{tcolorbox}[
  title=Finding 1: Atomic improves consistency across actors.,
  breakable,   
  fonttitle=\bfseries,
  enhanced,                        
  colback=my_lightblue!10,           %
  colbacktitle=my_lightblue,         %
  coltitle=black,                 %
  colframe=my_lightblue!80!black,    %
  coltext=black,                  %
  boxrule=0.5pt,
  arc=1mm
]
Compared with \texttt{BashOnly}, \texttt{Atomic} improves consistency
for all three actors, with the largest gain for
\texttt{Qwen3Coder-30B}. A plausible mechanism is that
\texttt{Atomic} reduces low-level environment-interaction errors,
which are more common in weaker actors and therefore leave more room
for improvement.
\end{tcolorbox}

\subsection{Exploration}

\begin{figure}[t]
    \centering
    \includegraphics[width=0.95\linewidth]{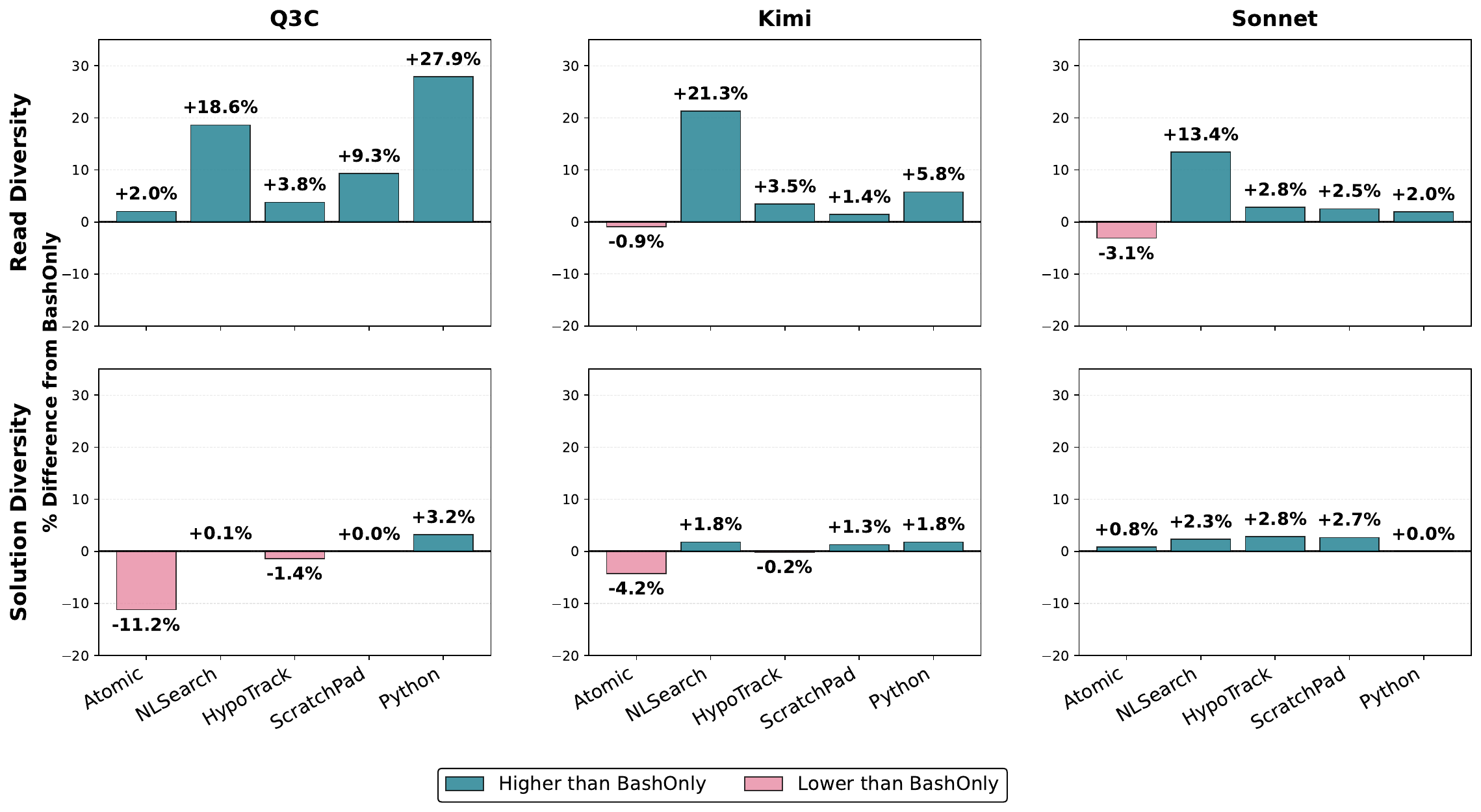}
    \caption{Effect of tool architecture on exploration and final-solution diversity. Bars show the difference from \texttt{BashOnly} for each actor--setup pair. The top row reports \emph{read diversity}, measured by pairwise Jaccard distance over files read across repeated attempts of the same problem instance. The bottom row reports \emph{solution diversity}, measured by pairwise CodeBLEU distance between generated patches. \texttt{NLSearch} is the only setup that consistently increases read diversity across all three actors, whereas effects on solution diversity are smaller and less uniform.}    
    \label{fig:diversity-diverging}
    \vspace{-5pt}
\end{figure}

Exploration matters because coding agents may under-explore
repository-level context~\citep{li2026contextbench}, and tool architectures that broaden
exploration may expose them to different code context and candidate
solutions. Figure~\ref{fig:diversity-diverging} shows that
\texttt{NLSearch} is the only setup that consistently increases
\emph{read diversity} across all three actors. The gains are substantial
for all models, whereas other setups have smaller or more actor-specific
effects. By contrast, changes in \emph{solution diversity} are much
smaller and less uniform. Across most actor--setup pairs, repeated
attempts under different tool architectures still converge to similar
final patches even when they inspect different parts of the repository.
Overall, tool architecture has a clearer effect on repository traversal
than on the diversity of final solutions.

To better understand this effect, we ask two follow-up questions: (1) what
drives the broader exploration under \texttt{NLSearch}, and (2) whether that
additional exploration is meaningful for the given problem instance.
Appendix~\ref{appdx:exploration-details} suggests that the broader
exploration is associated with more diverse early search queries under
the natural-language interface than under raw
\texttt{grep}/\texttt{find}-style search. It also shows that the
additional exploration improves recall of high-relevance files across
all three actors, though with lower precision than \texttt{BashOnly}.

\begin{tcolorbox}[
  title=\textbf{Finding 2: NLSearch improves exploration but not solution diversity.},
  breakable,
  fonttitle=\bfseries,
  enhanced,
  colback=my_lightblue!10,
  colbacktitle=my_lightblue,
  coltitle=black,
  colframe=my_lightblue!80!black,
  coltext=black,
  boxrule=0.5pt,
  arc=1mm
]
NLSearch consistently broadens repository coverage and increases recall of relevant files, but these gains only weakly translate to more diverse final solution strategies. The effect appears to come from more diverse early search queries, which broaden the code context seen during execution.
\end{tcolorbox}

\subsection{Efficiency}

\begin{figure}[t]
    \centering
    \includegraphics[width=0.95\linewidth]{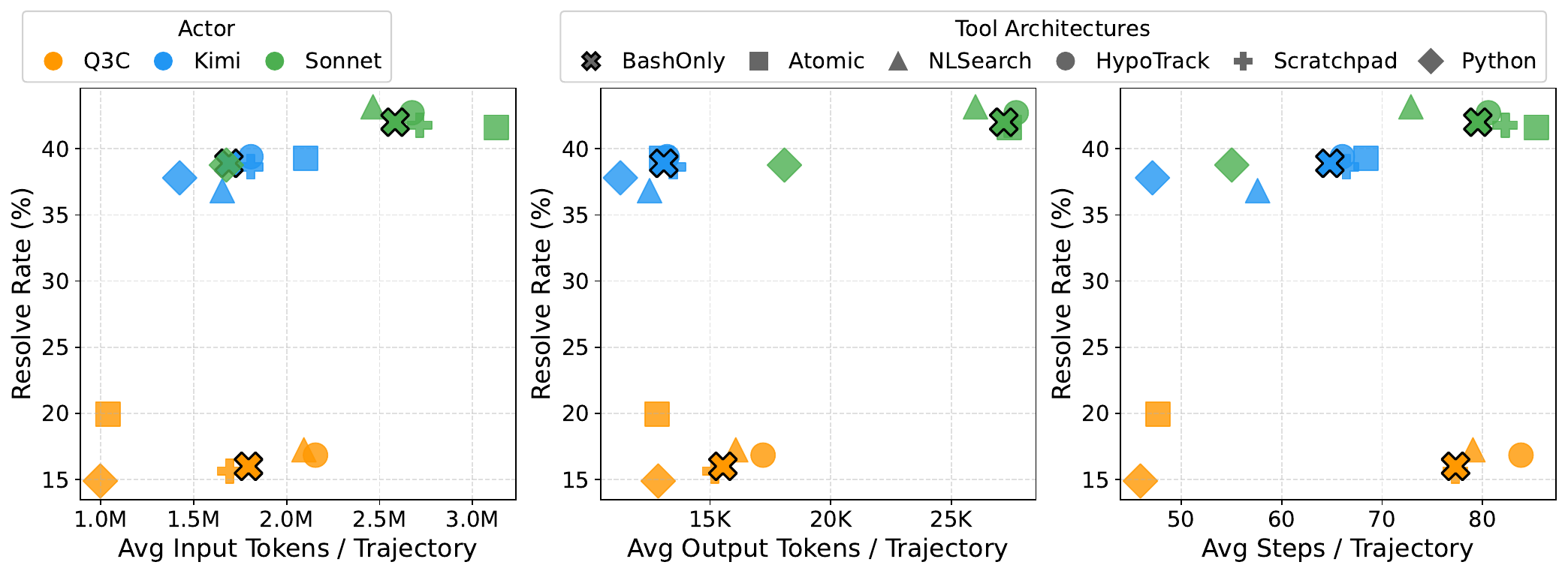}
    \caption{Efficiency comparison across tool architectures. Each point
represents one actor--setup pair; points closer to the upper-left of a
panel indicate a more favorable efficiency--performance trade-off,
achieving higher resolve rate with lower cost. The three panels show
resolve rate versus average input tokens, output tokens, and interaction
steps per trajectory. \texttt{Python} consistently achieves competitive
performance with fewer steps and lower token cost, whereas the
efficiency effect of \texttt{Atomic} depends on the actor: it reduces
cost for \texttt{Qwen3Coder-30B} but increases cost for
\texttt{Kimi-K2.5} and \texttt{Sonnet-4.5}.}
    \label{fig:eff-scatter}
    \vspace{-5pt}
\end{figure}

Besides consistency and exploration, the cost of a coding agent is also
important in deployment. Figure~\ref{fig:eff-scatter} shows a clear
overall pattern: \texttt{Python} is the most efficient setup across all
three actors. Relative to \texttt{BashOnly}, it achieves similar task
performance with substantially lower input-token cost and fewer
interaction steps, making it the clearest efficiency win in our study.
By contrast, the effect of \texttt{Atomic} is actor-dependent.
\texttt{Atomic} improves efficiency for \texttt{Qwen3Coder-30B}, but is
less efficient for the stronger actors, \texttt{Kimi-K2.5} and
\texttt{Sonnet-4.5}, where the increase in input tokens is especially
pronounced.

\textbf{Appendix~\ref{appdx:efficiency-details} suggests that these efficiency
differences mainly depend on how well each interface supports compound
interaction when tool capabilities are otherwise similar.} 
The \texttt{Python} interface encourages the actor to
bundle multiple operations into a single executable block, which helps
reduce both trajectory length and cumulative input-token cost. 
By contrast, the efficiency effect of \texttt{Atomic} depends on the
actor. For stronger actors, it can fragment work that would otherwise
be combined in bash into more turns, increasing overhead. For
\texttt{Qwen3Coder-30B}, this fragmentation is smaller, and
\texttt{Atomic} also reduces low-level interaction errors, which 
further improves efficiency. Appendix~\ref{appdx:efficiency-details}
provides the detailed analysis.

\begin{tcolorbox}[
  title=\textbf{Finding 3: Tool architectures supporting compound interaction are more efficient.},
  breakable,
  fonttitle=\bfseries,
  enhanced,
  colback=my_lightblue!10,
  colbacktitle=my_lightblue,
  coltitle=black,
  colframe=my_lightblue!80!black,
  coltext=black,
  boxrule=0.5pt,
  arc=1mm
]
\texttt{Python} consistently improves efficiency by allowing actors to
express more work in a single interaction. \texttt{Atomic}, by
contrast, is beneficial only when the actor already operates close to
an atomic style; otherwise, it fragments compound bash workflows into
more turns and raises cumulative token cost.
\end{tcolorbox}

\subsection{Why Cognitive Scaffolding Has Limited Effects}

The two cognitive-scaffolding setups, \texttt{Scratchpad} and
\texttt{HypoTrack}, have only limited effects in our main evaluation. We interpret this result as specific to
the lightweight scaffolds studied here, rather than as a claim about cognitive
scaffolding in general.
Appendix~\ref{appdx:cog-scaffold-analysis} suggests that, in our setting, actors usually
do not use these tools in ways that substantially change their reasoning. Under \texttt{Scratchpad},
actors often produce content that closely matches reasoning already
present in \texttt{BashOnly}. Under \texttt{HypoTrack}, actors rarely
maintain multiple competing hypotheses.

\begin{tcolorbox}[
  title=\textbf{Finding 4: Lightweight cognitive scaffolding has limited effect on actor reasoning.},
  breakable,
  fonttitle=\bfseries,
  enhanced,
  colback=my_lightblue!10,
  colbacktitle=my_lightblue,
  coltitle=black,
  colframe=my_lightblue!80!black,
  coltext=black,
  boxrule=0.5pt,
  arc=1mm
]
Although \texttt{Scratchpad} and \texttt{HypoTrack} provide explicit
interfaces for recording intermediate reasoning, they do not add retrieval, memory management,
new task information, or enforce a different reasoning policy. Actors therefore
mostly project their original reasoning patterns into these lightweight
scaffolds, suggesting that such scaffolds alone have limited effect in our
setting.
\end{tcolorbox}

\subsection{Generalization to Additional Coding Tasks}

We further validate whether the main tool-architecture effects extend beyond
our SWE-bench Live subset. Appendix~\ref{appendix:generalization-more-tasks}
reports results on additional coding tasks, including issue resolving, feature
implementation, and debugging. The results are directionally consistent with
the main findings: \texttt{Atomic} improves \(\consk{k}\), \texttt{NLSearch}
increases read diversity, and \texttt{Python} generally reduces input-token
cost and trajectory length.

\section{Related Work}

\noindent \textbf{Empirical studies of coding agents.}
Recent empirical work studies coding agents from several perspectives,
including task-specific analysis, real-world developer use,
planning behavior, and agent context or configuration
~\citep{ceka2025understandingsoftwareengineeringagents,
majgaonkar2025understandingcodeagentbehaviour,
liang2025swebenchillusionstateoftheartllms,
horikawa2025agenticrefactoringempiricalstudy,
watanabe2026useagenticcodingempirical,
wang2025empiricalstudyagentdeveloper,
liu2025empiricalstudyfailuresautomated,
galster2026configuringagenticaicoding,
chatlatanagulchai2025agentreadmesempiricalstudy}. These studies provide
insights about how coding agents are used in practice,
but they offer limited guidance on how the
organization and exposure of tools shape coding-agent behavior.

\noindent \textbf{Empirical studies of tool design.}
Another related line of work studies tool use more directly, especially
whether agents can retrieve and invoke the correct tool in large and
realistic tool ecosystems~\citep{wang2025mcpbenchbenchmarkingtoolusingllm,
fan2025mcptoolbenchlargescaleai}. 
However, it primarily evaluates whether an agent can use
the right tool rather than how tool architectures affect agent behavior.

\noindent \textbf{Agent non-functional properties.}
Recent work has also emphasized non-functional properties of agents,
including efficiency, robustness, and reliability
~\citep{gao2025lessempiricalstudyturncontrol,
wang2026agentnoisebenchbenchmarkingrobustnesstoolusing,
rabanser2026towards}. These studies motivate going beyond aggregate task
success, but they do not study which parts of the design
space produce these properties. 

\noindent \textbf{Tool design in coding agents.}
Many coding-agent systems propose concrete tool designs
~\citep{pan2026prometheuslonghorizoncodebasenavigation,yang2024swe,
liu2025reposcope,ouyang2024repograph,claudecode,
antoniades2024swe,openhands}. These systems directly inspire the tool
architectures we study. However, they typically introduce new tools as
part of end-to-end agent designs, changing capability and interface at
the same time. By contrast, our work isolates tool architecture itself
and studies its effect under roughly matched capabilities.

Please refer to Appendix~\ref{appdx:related} for detailed discussion of related work.

\section{Conclusion}
Our work isolates the effect of tool architecture in coding agents
under similar capabilities. We find that architectures with different
levels of abstraction can substantially affect consistency,
exploration, and efficiency, whereas cognitive-scaffolding
architectures have limited effect in our setting. These results
provide practical guidance for coding-agent tool design.

\section*{Acknowledgments}

We thank the anonymous reviewers for their valuable
comments and suggestions. This research was supported, in part by the Microsoft Internship Program, National Science Foundation (NSF)
Awards SHF-1901242, SHF-1910300, Proto-OKN 2333736,
IIS-2416835, DARPA VSPELLS - HR001120S0058, and ONR
N00014-23-1-2081. Any opinions, findings, and conclusions in this paper are those of the authors only and do not necessarily reflect the views of our sponsors.

\section*{LLM Usage Disclosure}

We used large language models to help polish the writing of the paper. The
authors reviewed and verified all content.

\bibliography{colm2026_conference}
\bibliographystyle{colm2026_conference}

\clearpage
\appendix
\section*{Appendix}

\section{Formal Definitions of Evaluation Metrics}
\label{appdx:metric-definitions}

We evaluate tool architectures along four complementary dimensions:
\emph{task resolve rate}, \emph{consistency}, \emph{exploration}, and
\emph{efficiency}.

\noindent \textbf{Task resolve rate.}
For benchmark instance \(i\) under setup \(s\), let \(c(i,s)\) denote
the number of successful attempts out of \(n\) total attempts. We define
\begin{equation}
\mathrm{ResolveRate}(i,s) = \frac{c(i,s)}{n}.
\end{equation}
Aggregating over instances gives the average probability that a
randomly sampled attempt resolves the problem instance under setup
\(s\).

\noindent \textbf{Consistency.}
Our main consistency metric is \consk{k}, which measures the
probability that \(k\) repeated attempts of the same problem instance
are all successful. For benchmark instance \(i\) under setup \(s\), let
\(c(i,s)\) denote the number of passing attempts out of \(n\) total
attempts. We define
\begin{equation}
\consk{k}(i,s) = \frac{\binom{c(i,s)}{k}}{\binom{n}{k}}.
\end{equation}
Equivalently, this is the empirical proportion of all size-\(k\)
subsets of repeated attempts that fully pass~\citep{chen2021evaluating}. Under i.i.d.\ repeated
attempts, it estimates the probability that \(k\) attempts would all
succeed. The metric equals \(1\) only when all attempts pass, and
decreases as failures appear among reruns. In particular, \consk{1}
reduces to the empirical pass rate, while larger \(k\) imposes a
stricter notion of repeatability by penalizing occasional failures more
strongly. In our experiments, we use \(k \in \{5,7,9\}\).

\noindent \textbf{Exploration.}
We quantify exploration from two complementary perspectives. The first
is \emph{file read diversity}, which measures how differently repeated
attempts traverse the repository. For each pair of attempts on the same
problem instance, we compute the Jaccard distance between their sets of
read files, and then average over attempt pairs and problem instances.
Let \(R_r(i,s)\) denote the set of files read in attempt \(r\) for
instance \(i\) under setup \(s\). We define
\begin{equation}
\mathrm{ReadDiv}(i,s)
=
\frac{1}{\binom{n}{2}}
\sum_{r_1<r_2}
\left(
1 -
\frac{|R_{r_1}(i,s)\cap R_{r_2}(i,s)|}
{|R_{r_1}(i,s)\cup R_{r_2}(i,s)|}
\right).
\end{equation}
Larger values indicate that repeated attempts tend to inspect more
different subsets of the repository.

The second exploration metric is \emph{solution diversity}. Even when
attempts read different files, they may still converge to nearly
identical outputs. We therefore measure the mean pairwise CodeBLEU
distance between generated solutions across repeated attempts of the
same problem instance. Let \(\mathrm{CodeBLEU}(y_{r_1}, y_{r_2})\)
denote the similarity between two generated solutions. We define
\begin{equation}
\mathrm{CodeBLEU\text{-}Dist}(i,s)
=
\frac{1}{\binom{n}{2}}
\sum_{r_1<r_2}
\left(1 - \mathrm{CodeBLEU}(y_{r_1}(i,s), y_{r_2}(i,s))\right).
\end{equation}
Larger values indicate that repeated attempts produce more diverse final
solutions.

\noindent \textbf{Efficiency.}
We measure efficiency at the trajectory level. A \emph{trajectory}
\(\tau\) is a complete interaction sequence for one attempt on one
benchmark instance:
\begin{equation}
\tau = (Q, A_1, O_1, A_2, O_2, \dots, A_T, O_T),
\end{equation}
where \(Q\) is the initial task prompt, \(A_t\) is the action issued by
the actor at step \(t\), \(O_t\) is the resulting environment
observation, and \(T\) is the total number of interaction steps before
termination.

The \emph{trajectory length} is the number of interaction steps:
\begin{equation}
\mathrm{Steps}(\tau) = T.
\end{equation}

We define \emph{input token cost} as the total number of tokens sent to
the model across all API calls in the trajectory. Because each API call
includes the original task together with the accumulated interaction
history, this quantity is cumulative across steps:
\begin{equation}
\mathrm{InputTok}(\tau)
=
\sum_{t=1}^{T}
\left|
\left(Q, A_1, O_1, \dots, A_{t-1}, O_{t-1}\right)
\right|.
\end{equation}

We define \emph{output token cost} as the total number of tokens
generated by the model across all steps:
\begin{equation}
\mathrm{OutputTok}(\tau) = \sum_{t=1}^{T} |A_t|.
\end{equation}

We additionally track \emph{observation tokens}, defined as the total
number of tokens returned by the environment across all steps:
\begin{equation}
\mathrm{ObsTok}(\tau) = \sum_{t=1}^{T} |O_t|.
\end{equation}

\begin{table}[t]
\centering
\caption{Atomic tools and their bash equivalents.}
\label{tab:atomic-tools}
\small
\begin{tabular}{lll}
\toprule
\textbf{Atomic tool} & \textbf{Purpose} &
\textbf{Equivalent bash command(s)} \\
\midrule
\texttt{search} & Search repository contents &
\texttt{grep -R}, \texttt{find}, \texttt{rg} \\
\texttt{view} & Read a bounded file region &
\texttt{sed -n}, \texttt{head}, \texttt{tail} \\
\texttt{str\_replace} & Replace a targeted string &
\texttt{sed -i}, \texttt{perl -0pi -e} \\
\texttt{create} & Create a new file &
\texttt{cat > file}, heredoc, \texttt{printf > file} \\
\texttt{bash} & Execute arbitrary shell commands &
\texttt{bash} \\
\bottomrule
\end{tabular}
\end{table}
\section{Additional Discussions for Evaluation Results}

\subsection{Performance}
\label{appdx:perf}

\begin{figure}[t]
    \centering
    \includegraphics[width=0.95\linewidth]{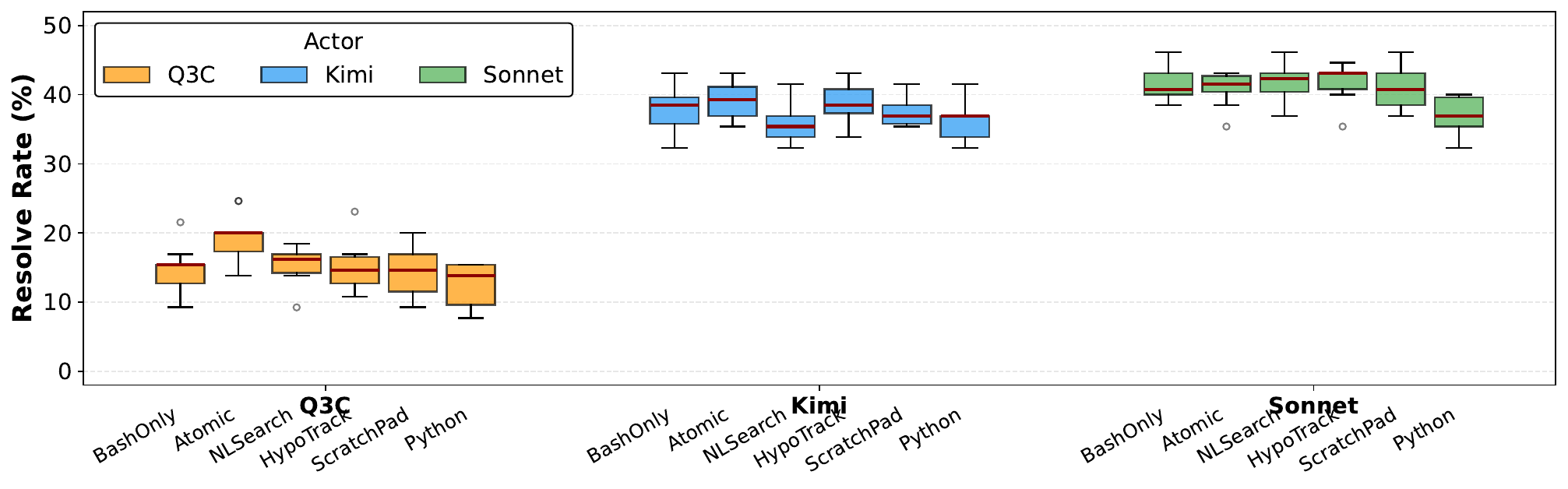}
    \caption{Overall resolve rates across actors and tool architectures.
Results are broadly similar across setups, reflecting our design choice
to keep tool capability close to the BashOnly baseline and vary
primarily the architecture of the interface.}
    \label{fig:overall-perf}
\end{figure}

Figure~\ref{fig:overall-perf} shows that overall performance remains in
a similar range across tool architectures. This is expected, since we
intentionally control tool capability to stay close to the BashOnly
setup and vary primarily the way those capabilities are organized and
exposed.

\subsection{Consistency Results with Additional Repeats}
\label{app:more-consistency-repeats}

Some of the consistency improvements from \texttt{Atomic} in the main
paper, especially for \texttt{Kimi-K2.5}, are numerically modest. To
check whether these gains remain stable under more repeated sampling, we
rerun the \texttt{BashOnly}--\texttt{Atomic} comparison with a larger
number of rollouts per instance and recompute \consk{5}, \consk{7}, and
\consk{9}.

Because this analysis substantially increases evaluation cost, we limit
it to the two open-weight actors: \texttt{Qwen3Coder-30B} and
\texttt{Kimi-K2.5}. In addition to the main-paper setting with 10
repeats, we also evaluate a 30-repeat variant while keeping the
benchmark instances and protocol otherwise unchanged.

Table~\ref{tab:consistency-more-repeats} shows that the main result is
stable. For \texttt{Qwen3Coder-30B}, \texttt{Atomic} improves all three
consistency metrics at both repeat counts, with gains of
$+0.059$, $+0.067$, and $+0.074$ at 10 repeats, and $+0.067$,
$+0.064$, and $+0.060$ at 30 repeats. For \texttt{Kimi-K2.5}, the
effect is smaller but remains uniformly positive, increasing from
$+0.014$, $+0.013$, and $+0.014$ at 10 repeats to $+0.015$,
$+0.015$, and $+0.016$ at 30 repeats. These results suggest that the consistency advantage of
\texttt{Atomic} is robust to substantially more repeated evaluation,
rather than a small-sample fluctuation from the 10-repeat setting.

\begin{table*}[t]
\centering
\small
\caption{Consistency robustness under additional repeats for open-weight actors.
We compare \texttt{BashOnly} and \texttt{Atomic} using 10 and 30 repeated
rollouts per instance. For each metric, $\Delta$ denotes
\texttt{Atomic} $-$ \texttt{BashOnly}.}
\label{tab:consistency-more-repeats}
\setlength{\tabcolsep}{3.4pt}
\renewcommand{\arraystretch}{1.15}
\begin{tabular}{lcccccccccc}
\toprule
\multirow{2}{*}{Actor} & \multirow{2}{*}{Repeats}
& \multicolumn{3}{c}{\consk{5}}
& \multicolumn{3}{c}{\consk{7}}
& \multicolumn{3}{c}{\consk{9}} \\
\cmidrule(lr){3-5} \cmidrule(lr){6-8} \cmidrule(lr){9-11}
& 
& Bash & Atomic & $\Delta$
& Bash & Atomic & $\Delta$
& Bash & Atomic & $\Delta$ \\
\midrule

\multirow{2}{*}{\texttt{Qwen3Coder-30B}}
& 10
& 0.046 & 0.106 & \textbf{+0.059}
& 0.031 & 0.097 & \textbf{+0.067}
& 0.020 & 0.094 & \textbf{+0.074} \\
& 30
& 0.044 & 0.111 & \textbf{+0.067}
& 0.033 & 0.097 & \textbf{+0.064}
& 0.027 & 0.087 & \textbf{+0.060} \\
\midrule

\multirow{2}{*}{\texttt{Kimi-K2.5}}
& 10
& 0.290 & 0.304 & \textbf{+0.014}
& 0.277 & 0.289 & \textbf{+0.013}
& 0.266 & 0.280 & \textbf{+0.014} \\
& 30
& 0.303 & 0.318 & \textbf{+0.015}
& 0.292 & 0.307 & \textbf{+0.015}
& 0.283 & 0.300 & \textbf{+0.016} \\
\bottomrule
\end{tabular}
\end{table*}

\subsection{Mechanism Analysis: Atomic Reduces Environment-Interaction
Errors}
\label{appdx:consistency-mechanism}
To help explain why \texttt{Atomic} yields the clearest consistency
gains, we examine environment-interaction errors as a possible
mechanism. This analysis is motivated by prior work in software
dependability~\cite{avizienis2004basic}, which analyzes unreliability
through a fault--error--failure chain, and by recent work on agent
reliability~\cite{gupta2026reliabilitybench,rabanser2026towards},
which studies repeated execution together with robustness to tool and
API failures. Following this perspective, we treat
environment-interaction errors as a plausible lens for understanding why
some tool architectures lead to lower repeated-run reliability than
others.

\begin{figure}
    \centering
    \includegraphics[width=0.9\linewidth]{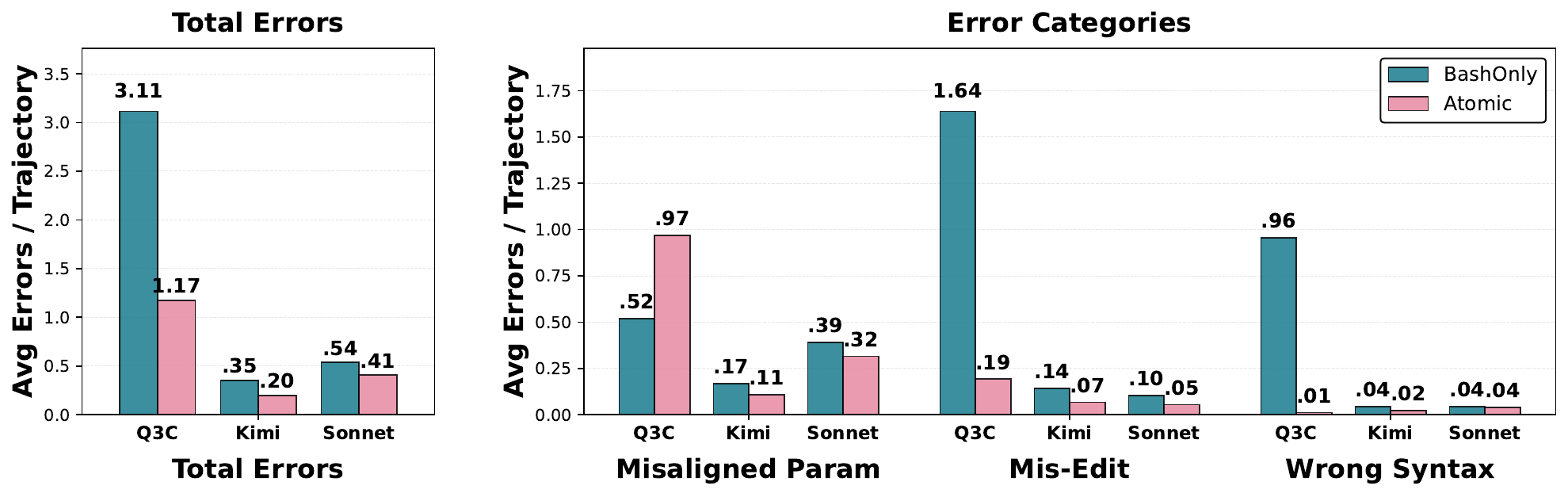}
    \caption{\texttt{Atomic} is associated with fewer
environment-interaction errors, especially for weaker actors. The
figure shows the average number of environment-interaction errors per
trajectory under \texttt{BashOnly} and \texttt{Atomic}, both in total
and broken down by error category for each actor. The largest reduction
appears for \texttt{Qwen3Coder-30B}, which also benefits most from
\texttt{Atomic} in consistency.}
    \label{fig:atomic-error-analysis}
\end{figure}

We focus on the comparison between \texttt{BashOnly} and
\texttt{Atomic}, because \texttt{Atomic} is the only setup that
improves \consk{k} for all three actors and yields the largest gains for
\texttt{Qwen3Coder-30B}. Figure~\ref{fig:atomic-error-analysis} shows that these
consistency differences are accompanied by corresponding differences in
environment-interaction errors. Under \texttt{BashOnly},
\texttt{Qwen3Coder-30B} incurs 3.11 interaction errors per trajectory
on average, compared with 0.35 for \texttt{Kimi-K2.5} and 0.54 for
\texttt{Sonnet-4.5}. This ordering mirrors the consistency results in
Table~\ref{tab:consistency-all-actors}: the actor with the lowest
\consk{k} values also exhibits the highest baseline rate of
environment-interaction errors.

To examine this pattern more closely, we divide
environment-interaction errors into three categories (the ``Error Categories'' subplot in Figure~\ref{fig:atomic-error-analysis}).
\emph{Misaligned-param} refers to cases where the agent applies a tool
to the wrong target, such as a nonexistent file path, an incorrect
working directory, or an invalid runtime target. \emph{Mis-edit} refers
to editing mistakes that leave the program in a broken state, such as
introducing \texttt{SyntaxError}, \texttt{IndentationError}, or other
execution-breaking edit failures. \emph{Wrong-syntax} refers to
malformed tool invocations, such as invalid shell, \texttt{sed}, or
\texttt{awk} expressions, or an incorrect binary name.

The category breakdown suggests that the reduction is concentrated in
the error types most directly targeted by a structured interface. For
\texttt{Qwen3Coder-30B}, \texttt{Atomic} reduces \emph{mis-edit}
errors from 1.64 to 0.19 and \emph{wrong-syntax} errors from 0.96 to
0.01. These are precisely the failure modes that should become less
common when free-form shell editing and command construction are
replaced by narrower, constrained operations. For \texttt{Kimi-K2.5}
and \texttt{Sonnet-4.5}, the corresponding reductions are smaller,
largely because these error categories are already rare under
\texttt{BashOnly}. This is consistent with the smaller consistency gains
of \texttt{Atomic} for the stronger actors.

One exception is \emph{misaligned-param}. For
\texttt{Qwen3Coder-30B}, this category increases under
\texttt{Atomic}, from 0.52 to 0.97. A plausible explanation is that
\texttt{Atomic} enforces stricter parameter validation than
unconstrained bash commands. Under this interpretation, some mistakes
that might previously have failed silently, partially applied, or gone
unnoticed under \texttt{BashOnly} are instead surfaced explicitly as
parameter-alignment errors. The increase therefore does not necessarily
imply worse high-level localization; it may instead reflect more
explicit detection and typing of the same underlying mistakes. Even with
this increase, the large reductions in \emph{mis-edit} and
\emph{wrong-syntax} dominate the overall change in Qwen's total error
count.

Taken together, the error analysis is consistent with the interpretation
that \texttt{Atomic}'s consistency gains arise in part from reducing
low-level environment-interaction errors, especially for actors that are
more prone to such errors under \texttt{BashOnly}. Note that
this analysis does not directly test the causal link
between environment-interaction errors and task-level inconsistency. Instead, we
follow prior work in software dependability and agent reliability in
treating such errors as a plausible explanatory lens for repeated-run
reliability.

\subsection{Why NLSearch Broadens Exploration and What It Covers}
\label{appdx:exploration-details}

\noindent\textbf{NLSearch produces the strongest gains in repository
exploration, while effects on downstream solution diversity remain
limited.}
Figure~\ref{fig:diversity-diverging} in the main paper shows that \texttt{NLSearch} is
the only setup that consistently increases \emph{read diversity} across
all three actors. The gains are substantial for all models, whereas
most other setups have smaller or more actor-specific effects. By
contrast, \emph{solution diversity} changes much less. Across most
actor--setup pairs, repeated attempts under different tool
architectures still converge to similar final patches even when they
inspect different parts of the repository. Thus, the effect of tool
architecture is more pronounced on repository traversal than on
downstream solution behavior.

A notable exception is \texttt{Atomic} for \texttt{Qwen3Coder-30B},
which sharply reduces solution diversity. This aligns with the
consistency analysis (Section~\ref{sec:eval:consistency}):
\texttt{Atomic} substantially reduces Qwen's \emph{mis-edit} and
\emph{wrong-syntax} errors, removing a major source of accidental patch
variation. With less broken or noisy editing, repeated attempts
converge to a narrower set of cleaner final patches.

\noindent\textbf{A likely reason is that NLSearch induces more diverse
early search actions.}
To better understand why \texttt{NLSearch} broadens repository
exploration, we next examine how search is performed early in a
trajectory. It is worth noting that \texttt{NLSearch} is implemented as
a search sub-agent using the same base LLM as the actor. We do not rely
on additional program-analysis structures~\citep{liu2025reposcope,ouyang2024repograph,ma2024understand} or semantic embeddings~\citep{augcodectx} to
support natural-language search. As a result, the observed effects are
best understood as coming from how the search interface is provided,
rather than from additional underlying information. Qualitatively,
actors under \texttt{BashOnly} often repeat similar
\texttt{grep}/\texttt{find} patterns across repeated attempts, whereas
under \texttt{NLSearch} they issue more varied natural-language
queries.

To quantify this effect, we compare \texttt{BashOnly} and
\texttt{NLSearch} during early exploration. We extract search actions
from the first 20 steps of each rollout as a proxy for initial
repository search. For \texttt{NLSearch}, these actions are
natural-language retrieval queries; for \texttt{BashOnly}, they are the
search strings passed to \texttt{grep} and \texttt{find}. Table~
\ref{tab:search-action-diversity} shows that \texttt{NLSearch} yields
higher diversity under both Jaccard distance and normalized
Levenshtein distance for all three actors. This suggests that
\texttt{NLSearch} broadens repository traversal by inducing more
diverse search formulations early in the trajectory.

\begin{table}[t]
\caption{Diversity of early search actions under \texttt{NLSearch} and \texttt{BashOnly}. We extract search actions from the first 20 steps of each rollout and compare them across repeated runs of the same benchmark instance. For \texttt{NLSearch}, actions are natural-language retrieval queries; for \texttt{BashOnly}, they are search patterns issued through \texttt{grep} and \texttt{find}. We report Jaccard distance and normalized Levenshtein distance over these strings. Higher values indicate more diverse search formulations.}
\label{tab:search-action-diversity}
\centering
\small
\begin{tabular}{llcc}
\toprule
Actor & Setup & Jaccard & Levenshtein \\
\midrule
\multirow{2}{*}{\centering \texttt{Qwen3Coder-30B}}
& NLSearch & 0.850 & 0.579 \\
& BashOnly  & 0.692 & 0.486 \\
\midrule
\multirow{2}{*}{\centering \texttt{Kimi-K2.5}}
& NLSearch & 0.873 & 0.563 \\
& BashOnly  & 0.634 & 0.451 \\
\midrule
\multirow{2}{*}{\centering \texttt{Sonnet-4.5}}
& NLSearch & 0.862 & 0.521 \\
& BashOnly  & 0.681 & 0.447 \\
\bottomrule
\end{tabular}
\end{table}

\begin{table}[t]
\centering
\caption{Manual validation of the high-relevant-file proxy. We compare
the proxy label against human judgments of whether each accessed file is
helpful for resolving the issue.}
\label{tab:manual-high-relevant-validation}
\begin{tabular}{lcc}
\toprule
 & Human Helpful & Human Not Helpful \\
\midrule
Proxy Relevant     & 561 & 33 \\
Proxy Not Relevant & 14  & 11 \\
\bottomrule
\end{tabular}
\end{table}

\begin{table}[t]
\centering
\small
\caption{Quality of explored context under different tool architectures. For each actor--setup pair, we compare the files read during a rollout against a conservative set of \emph{high-relevant} files for that benchmark instance. A file is labeled high-relevant if it is read by at least three successful trajectories from every setup. Values in parentheses show the difference relative to \texttt{BashOnly} for the same actor. \texttt{NLSearch} consistently improves recall, but at the cost of lower precision.}
\setlength{\tabcolsep}{6pt}
\renewcommand{\arraystretch}{1.2}
\begin{tabular}{llcc}
\toprule
\textbf{Actor} & \textbf{Setup} & \textbf{Precision} & \textbf{Recall} \\
\midrule
\multirow{6}{*}{\texttt{Qwen3Coder-30B}}
& BashOnly   & 0.938           & 0.406           \\
& Atomic     & \textbf{0.944 (+0.006)}  & 0.431 (+0.025)  \\
& NLSearch  & 0.891 (-0.046)  & \textbf{0.452 (+0.046)}  \\
& HypoTrack  & 0.938 (+0.001)  & 0.418 (+0.012)  \\
& ScratchPad & 0.932 (-0.006)  & 0.410 (+0.004)  \\
& Python     & 0.888 (-0.050)  & 0.365 (-0.042)  \\
\midrule
\multirow{6}{*}{\texttt{Kimi-K2.5}}
& BashOnly   & 0.939           & 0.548           \\
& Atomic     & 0.902 (-0.037)  & 0.571 (+0.023)  \\
& NLSearch  & 0.844 (-0.095)  & \textbf{0.601 (+0.053)}  \\
& HypoTrack  & \textbf{0.950 (+0.011)}  & 0.555 (+0.007)  \\
& ScratchPad & 0.946 (+0.007)  & 0.549 (+0.001)  \\
& Python     & 0.937 (-0.002)  & 0.529 (-0.019)  \\
\midrule
\multirow{6}{*}{\texttt{Sonnet-4.5}}
& BashOnly   & 0.919           & 0.578           \\
& Atomic     & 0.915 (-0.004)  & 0.588 (+0.010)  \\
& NLSearch  & 0.816 (-0.104)  & \textbf{0.643 (+0.064)}  \\
& HypoTrack  & 0.927 (+0.008)  & 0.563 (-0.016)  \\
& ScratchPad & 0.932 (+0.013)  & 0.564 (-0.014)  \\
& Python     & \textbf{0.946 (+0.027)}  & 0.532 (-0.047)  \\
\bottomrule
\end{tabular}
\label{tab:retrieval-precision-recall}
\end{table}

\noindent\textbf{The additional exploration induced by NLSearch also
covers more relevant files, but with lower precision.}
Broader exploration is only useful if the additional files accessed are
actually relevant to the problem instance. To assess this, we introduce
a conservative file-level proxy. For each benchmark instance, a file is
labeled \emph{high-relevant} if it is read by at least three successful
trajectories from \emph{every} setup. This definition focuses on files
that recur across successful resolutions under diverse tool
architectures, while reducing bias toward any single setup.

\paragraph{Manual validation of the high-relevant-file proxy.}
Because the high-relevant-file set is a proxy rather than a human-labeled
gold standard, we additionally validate it with a small manual annotation
study. We sample 30 trajectories and label 622 file operations according to
whether the accessed file is helpful for resolving the corresponding issue.
The annotation protocol is LLM-assisted: we first use Codex to generate
reference reasoning for each file access, and then ask human annotators to
judge the final helpfulness label. We exclude 3 cases where the human
judgment and the reference reasoning diverge, leaving 619 file operations
for evaluation.

Table~\ref{tab:manual-high-relevant-validation} shows that the
high-relevant-file proxy aligns well with human helpfulness judgments in
this sampled validation set. The proxy achieves 94.4\% precision and
97.6\% recall:
\[
\mathrm{Precision} = \frac{561}{561 + 33} = 94.4\%, \qquad
\mathrm{Recall} = \frac{561}{561 + 14} = 97.6\%.
\]
This result supports the use of the proxy as a conservative approximation
to helpful file access. In particular, it suggests that the proxy is not
merely capturing files that are frequently read under broader-exploration
setups such as NLSearch, but is generally aligned with files humans judge
as useful for resolving the task. At the same time, we treat this validation
as supporting evidence rather than a perfect ground truth, since file
usefulness can depend on trajectory context and may include files that help
reasoning without appearing in the final patch.

\noindent\textbf{Table~\ref{tab:retrieval-precision-recall} evaluates the files read
under each setup against this high-relevance set. } \texttt{NLSearch}
achieves the highest recall for all three actors, indicating that it
covers more files plausibly needed for successful resolution. At the
same time, its precision is lower than the \texttt{BashOnly} baseline.
In other words, \texttt{NLSearch} broadens exploration in a meaningful
sense by improving coverage of relevant files, but it also introduces
additional context that the actor must filter and use effectively.

\subsection{Detailed Efficiency Analysis and Compound Interaction}
\label{appdx:efficiency-details}

\begin{figure}
    \centering
    \includegraphics[width=0.9\linewidth]{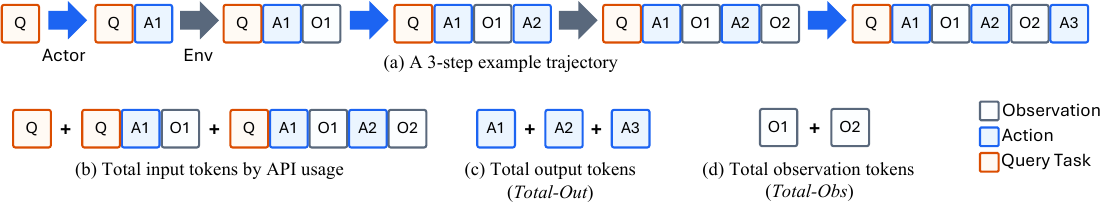}
\caption{Illustration of the token metrics used in
Table~\ref{tab:py-efficiency}. The decomposition clarifies why trajectory length strongly affects cumulative input cost: each additional interaction step requires
another API call, which includes both the new action and
the cumulative interaction history. Panel (a) shows a three-step trajectory
with an initial query task $Q$, actions $A_t$, and observations $O_t$.
Panel (b) illustrates cumulative input tokens as the total tokens sent
to the model across API calls, including the repeated query context and
the growing history of prior actions and observations. Panel (c) shows
total output tokens, defined as the sum of all action tokens generated
by the model. Panel (d) shows total observation tokens, defined as the
sum of all environment outputs returned to the model. ~\looseness=-1}
    \label{fig:tok-cost}
\end{figure}

\begin{table}[t]
\caption{Efficiency comparison between \texttt{BashOnly} and
\texttt{Python}. Each cell reports the mean value of the corresponding
trajectory-level metric.}
\label{tab:py-efficiency}
\centering
\small
\begin{tabular}{llccccc}
\toprule
Actor & Setup & Steps & Total-Out & Total-Obs & Out/Step & Obs/Step \\
\midrule
\multirow{2}{*}{\centering \texttt{Qwen3Coder-30B}}
& \texttt{BashOnly} & 77 & 15{,}535 & 15{,}228 & 197 & 230 \\
& \texttt{Python}   & 46 & 12{,}854 & 15{,}992 & 315 & 410 \\
\midrule
\multirow{2}{*}{\centering \texttt{Kimi-K2.5}}
& \texttt{BashOnly} & 65 & 13{,}087 & 23{,}670 & 196 & 389 \\
& \texttt{Python}   & 47 & 11{,}288 & 26{,}298 & 226 & 607 \\
\midrule
\multirow{2}{*}{\centering \texttt{Sonnet-4.5}}
& \texttt{BashOnly} & 80 & 27{,}168 & 32{,}707 & 355 & 438 \\
& \texttt{Python}   & 55 & 18{,}082 & 28{,}915 & 334 & 561 \\
\bottomrule
\end{tabular}
\end{table}

\begin{figure}[t]
    \centering
    \includegraphics[width=0.5\linewidth]{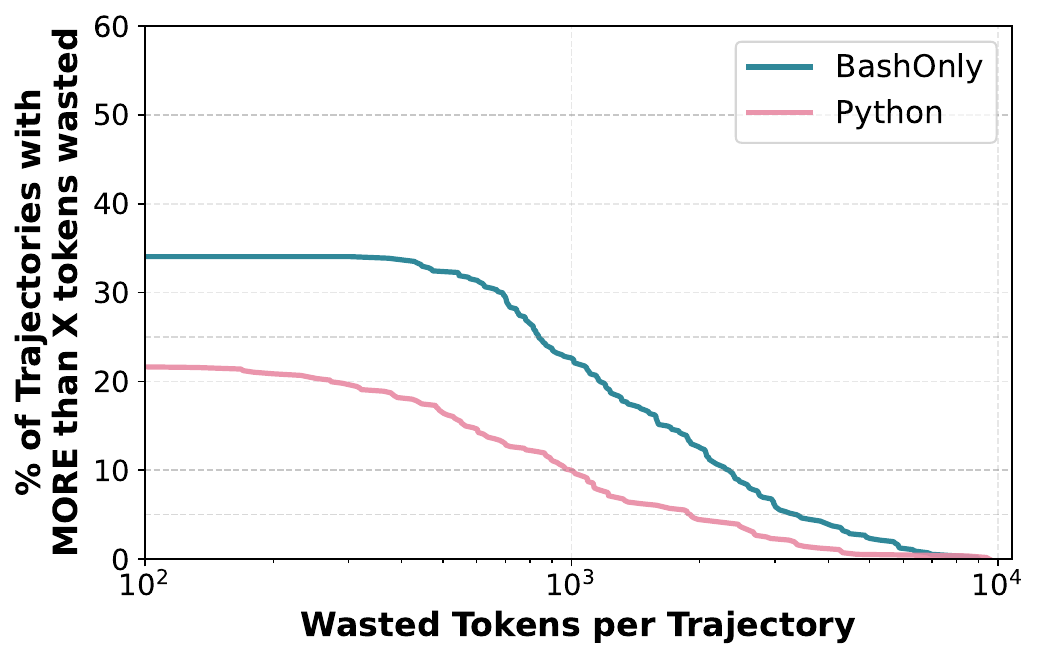}
    \caption{Wasted tokens from speculative revision under
\texttt{BashOnly} and \texttt{Python} for Sonnet-4.5. For each trajectory, we measure
tokens spent on consecutive writes to the same file before a validating
action. The complementary cumulative distribution shows that
\texttt{Python} substantially reduces these wasted tokens, especially
in the high-waste tail.}
    \label{fig:consec-write}
\end{figure}

\noindent\textbf{The efficiency advantage of \texttt{Python} mainly
comes from reducing interaction steps.}
Compared with \texttt{BashOnly}, \texttt{Python} substantially shortens
trajectories for all three actors. At the same time, its total output
tokens and total observation tokens remain broadly similar rather than
uniformly smaller. This suggests that the main efficiency gain does not
come from making each interaction cheaper, but from reducing the number
of API calls needed to complete an attempt.

Figure~\ref{fig:tok-cost} clarifies this distinction. Total input tokens
accumulate across API calls, so longer trajectories repeatedly pay for
the query context together with a growing history of prior actions and
observations. By contrast, total output tokens and total observation
tokens measure the amount of generated action content and returned
environment output, respectively. As a result, an interface can reduce
overall cost mainly by shortening the trajectory, even if the amount of
work performed in each individual step stays similar or becomes larger.

Table~\ref{tab:py-efficiency} is consistent with this explanation.
Across all three actors, \texttt{Python} reduces interaction steps
substantially relative to \texttt{BashOnly}: from 77 to 46 for
\texttt{Qwen3Coder-30B}, from 65 to 47 for \texttt{Kimi-K2.5}, and from
80 to 55 for \texttt{Sonnet-4.5}. For \texttt{Qwen3Coder-30B} and
\texttt{Kimi-K2.5}, total output and total observation tokens remain
broadly similar under the two setups, while per-step output and
observation tokens are often larger under \texttt{Python}. This
indicates that the actor is carrying out more work in each interaction.
The resulting efficiency gain therefore comes primarily from step
reduction and the associated reduction in cumulative input cost. 

\noindent\textbf{For \texttt{Sonnet-4.5}, the efficiency gain of
\texttt{Python} comes not only from fewer steps, but also from
reducing speculative full-file rewrites through faster execution
feedback.}
With \texttt{BashOnly}, we often observe consecutive full-file rewrites to
the same file, where the agent repeatedly revises content before
executing or validating it. This resembles a form of \textit{speculative
revision}: the agent keeps rewriting the file without receiving feedback
that would guide the next change. Under \texttt{Python}, this behavior
is much less frequent, likely because editing and execution are more
naturally coupled within a single interaction, so the agent receives
faster feedback and makes more directed modifications.

Figure~\ref{fig:consec-write} illustrates this effect. For each
trajectory, we measure the number of tokens spent on consecutive writes
to the same file before a validating action, and plot the complementary
cumulative distribution across trajectories. The curve for
\texttt{Python} lies consistently below that of \texttt{BashOnly},
showing that \texttt{Python} produces fewer trajectories with large
amounts of waste from speculative revision. This reduction is
especially pronounced in the high-waste tail, suggesting that part of
\texttt{Python}'s efficiency gain for \texttt{Sonnet-4.5} comes from
avoiding repeated unguided rewrites rather than only shortening the
number of steps.

\noindent\textbf{The effect of \texttt{Atomic} depends on how compound
the actor's original bash interaction is.}
If an actor already combines multiple low-level operations into a single
bash step, then replacing bash with \texttt{Atomic} can split that work
across more turns and increase overhead. If the actor is already closer
to an atomic interaction style under \texttt{BashOnly}, the same
interface imposes less fragmentation and may even improve efficiency.

Table~\ref{tab:atomic-effects} shows that this ordering closely matches
the efficiency effect of replacing \texttt{BashOnly} with
\texttt{Atomic}. For \texttt{Sonnet-4.5} and \texttt{Kimi-K2.5}, which
show more compound bash behavior, \texttt{Atomic} increases both steps
and input tokens. By contrast, \texttt{Qwen3Coder-30B}, which is closer
to an atomic interaction style under \texttt{BashOnly}, becomes more
efficient under \texttt{Atomic}. For \texttt{Qwen3Coder-30B}, however,
the benefit of reducing low-level interaction errors appears to
outweigh the overhead of a more fine-grained interface; see Appendix~
\ref{appdx:consistency-mechanism}. This pattern suggests that the
efficiency effect of tool architecture depends not only on what
operations the interface makes available, but also on whether its
granularity matches the actor's natural interaction style.

\begin{table}[t]
\caption{Estimated compounding of \texttt{BashOnly} trajectories,
measured by parsing each bash step into atomic-style operations. Higher
\texttt{Ops/Step} indicates that a single bash step combines more
underlying operations.}
\label{tab:atomic-compound-baseline}
\centering
\small
\setlength{\tabcolsep}{5pt}
\begin{tabular}{lcccccccc}
\toprule
Actor & READ & SEARCH & EDIT & WRITE & RUN & GIT & MISC & Ops/Step \\
\midrule
\texttt{Qwen3Coder-30B} & 18.1 & 14.1 & 7.6 & 5.3  & 17.2 & 2.0 & 24.1 & 1.154 \\
\texttt{Kimi-K2.5}      & 30.0 & 10.9 & 0.5 & 8.2  & 20.6 & 2.5 & 11.1 & 1.403 \\
\texttt{Sonnet-4.5}     & 32.3 & 18.6 & 0.5 & 18.1 & 30.5 & 3.4 & 12.3 & 1.500 \\
\bottomrule
\end{tabular}
\end{table}

\begin{table}[t]
\caption{Effect of replacing \texttt{BashOnly} with \texttt{Atomic}.
Actors with more compound bash behavior tend to incur more step and
token overhead under atomic tools.}
\label{tab:atomic-effects}
\centering
\small
\setlength{\tabcolsep}{5pt}
\begin{tabular}{l c >{\centering\arraybackslash}p{2.2cm}
>{\centering\arraybackslash}p{3.0cm}
>{\centering\arraybackslash}p{3.0cm}}
\toprule
Actor & Ops/Step & Steps & Input Tokens & Output Tokens \\
& & (bash$\to$atomic) & (bash$\to$atomic) & (bash$\to$atomic) \\
\midrule
\texttt{Qwen3Coder-30B}
& 1.15
& 64 $\rightarrow$ 43\down{33\%}
& 980K $\rightarrow$ 662K\down{32\%}
& 11444 $\rightarrow$ 9672\down{15\%} \\

\texttt{Kimi-K2.5}
& 1.40
& 56 $\rightarrow$ 59\up{5\%}
& 950K $\rightarrow$ 1141K\up{20\%}
& 9556 $\rightarrow$ 9659\up{1\%} \\

\texttt{Sonnet-4.5}
& 1.50
& 74 $\rightarrow$ 80\up{8\%}
& 2140K $\rightarrow$ 2576K\up{20\%}
& 25357 $\rightarrow$ 25089\down{1\%} \\
\bottomrule
\end{tabular}
\end{table}

\paragraph{Python does not introduce new action capabilities.}
Finally, we manually inspect 100 Python actions sampled from different actors
and trajectories to check whether the Python setup changes the available
actions, rather than only the interaction style. For each action, we identify
whether the same operation can be carried out in BashOnly, with Codex assisting
the mapping. Table~\ref{tab:python-bash-counterparts} summarizes the result.

\begin{table}[t]
\centering
\caption{Manual inspection of whether Python actions introduce capabilities
beyond BashOnly.}
\label{tab:python-bash-counterparts}
\begin{tabular}{rl}
\toprule
Count & BashOnly counterpart \\
\midrule
89 & Directly correspond to operations available in \texttt{BashOnly} \\
8  & Python test/check scripts that \texttt{BashOnly} commonly create and run from bash \\
3  & Broken-syntax Python code whose intended operation is unclear \\
\bottomrule
\end{tabular}
\end{table}

\begin{table}[t]
\centering
\caption{Effect of \texttt{Atomic} on consistency across coding tasks. Each cell
reports \(\Delta \consk{5}\), computed as \texttt{Atomic} minus
\texttt{BashOnly}, in percentage points. Positive values indicate higher
consistency under \texttt{Atomic}.}
\label{tab:generalization-pass5}
\resizebox{\linewidth}{!}{
\begin{tabular}{lrrrrr}
\toprule
Actor & SWE-bench Live & SWE-bench Verified & Feature Impl. & Debugging & Overall \\
\midrule
\texttt{Qwen3Coder-30B} & +5.9 & +15.6 & +8.4 & +8.0 & +9.5 \\
\texttt{Kimi-K2.5}      & +1.4 & +4.4  & +4.2 & +0.0 & +2.5 \\
\bottomrule
\end{tabular}
}
\end{table}

\begin{table}[t]
\centering
\caption{Effect of \texttt{NLSearch} on exploration across coding tasks. Each
cell reports the relative change in read diversity, computed as
\texttt{NLSearch} relative to \texttt{BashOnly}. Positive values indicate higher
read diversity under \texttt{NLSearch}.}
\label{tab:generalization-read-diversity}
\resizebox{\linewidth}{!}{
\begin{tabular}{lrrrrr}
\toprule
Actor & SWE-bench Live & SWE-bench Verified & Feature Impl. & Debugging & Overall \\
\midrule
\texttt{Qwen3Coder-30B} & +18.6\% & +23.3\% & +2.7\% & +7.5\% & +13.0\% \\
\texttt{Kimi-K2.5}      & +21.3\% & +13.0\% & +0.6\% & +2.7\% & +9.4\% \\
\bottomrule
\end{tabular}
}
\end{table}

\begin{table}[t]
\centering
\caption{Effect of \texttt{Python} on input-token efficiency across coding
tasks. Each cell reports \(\Delta\) input tokens, computed as \texttt{Python}
minus \texttt{BashOnly}, in millions. Negative values indicate lower input-token
cost under \texttt{Python}.}
\label{tab:generalization-input-tokens}
\resizebox{\linewidth}{!}{
\begin{tabular}{lrrrrr}
\toprule
Actor & SWE-bench Live & SWE-bench Verified & Feature Impl. & Debugging & Overall \\
\midrule
\texttt{Qwen3Coder-30B} & -0.8 & -0.6 & -1.1 & +0.2 & -0.6 \\
\texttt{Kimi-K2.5}      & -0.3 & +0.0 & -1.1 & -0.2 & -0.4 \\
\bottomrule
\end{tabular}
}
\end{table}

\begin{table}[t]
\centering
\caption{Effect of \texttt{Python} on step efficiency across coding tasks. Each
cell reports \(\Delta\) steps, computed as \texttt{Python} minus
\texttt{BashOnly}. Negative values indicate shorter trajectories under
\texttt{Python}.}
\label{tab:generalization-steps}
\resizebox{\linewidth}{!}{
\begin{tabular}{lrrrrr}
\toprule
Actor & SWE-bench Live & SWE-bench Verified & Feature Impl. & Debugging & Overall \\
\midrule
\texttt{Qwen3Coder-30B} & -31.4 & -25.0 & -40.3 & -6.9 & -25.9 \\
\texttt{Kimi-K2.5}      & -17.7 & -8.7  & -25.0 & -8.8 & -15.1 \\
\bottomrule
\end{tabular}
}
\end{table}

Overall, 97 of the 100 inspected Python actions correspond to operations already
available in BashOnly. Note that the 8 test/check scripts do not add new action
capabilities, since
\texttt{BashOnly} actors commonly create the same script with a heredoc and
execute it with \texttt{python3}. This supports our interpretation that
Python improves efficiency by changing how actors express compound interactions,
rather than by introducing new repository information or new actions.

\subsection{How Tool-Architecture Effects Generalize to More Coding Tasks}
\label{appendix:generalization-more-tasks}

We study how the main tool-architecture effects generalize beyond our
SWE-bench Live~\citep{zhang2025swe} subset by evaluating additional coding tasks on 72 instances
from 23 repositories across issue resolving, feature implementation, and
debugging. Specifically, we use SWE-bench Verified~\citep{jimenez2023swe} for issue resolving and
SWE-bench Pro~\citep{deng2025swebenchproaiagents} for feature implementation. For debugging, we construct a separate
SWE-bench split by selecting instances that contain full stack traces in the
issue description.

Due to budget constraints, we run this study with the two open-weight actors,
\texttt{Qwen3Coder-30B} and \texttt{Kimi-K2.5}. We do not repeat the full main
evaluation. Instead, this validation focuses on the pairwise comparisons most
directly tied to our main findings: \texttt{Atomic} vs. \texttt{BashOnly} for
consistency, \texttt{NLSearch} vs. \texttt{BashOnly} for exploration, and
\texttt{Python} vs. \texttt{BashOnly} for efficiency. For each actor, task, and
setup, we run 5 independent rollouts per instance. Tables~\ref{tab:generalization-pass5}--\ref{tab:generalization-steps} report the results.

Overall, the additional validation supports the main findings beyond the
original SWE-bench Live subset. Table~\ref{tab:generalization-pass5} shows that
\texttt{Atomic} improves \(\consk{5}\) for both actors overall.
Table~\ref{tab:generalization-read-diversity} shows that \texttt{NLSearch}
increases read diversity across tasks. Tables~\ref{tab:generalization-input-tokens}
and~\ref{tab:generalization-steps} show that \texttt{Python} generally reduces
both input-token cost and trajectory length.

\section{Why Cognitive Scaffolding Tools Have Limited Effects}
\label{appdx:cog-scaffold-analysis}

\begin{figure}[t]
    \centering
    \includegraphics[width=0.75\linewidth]{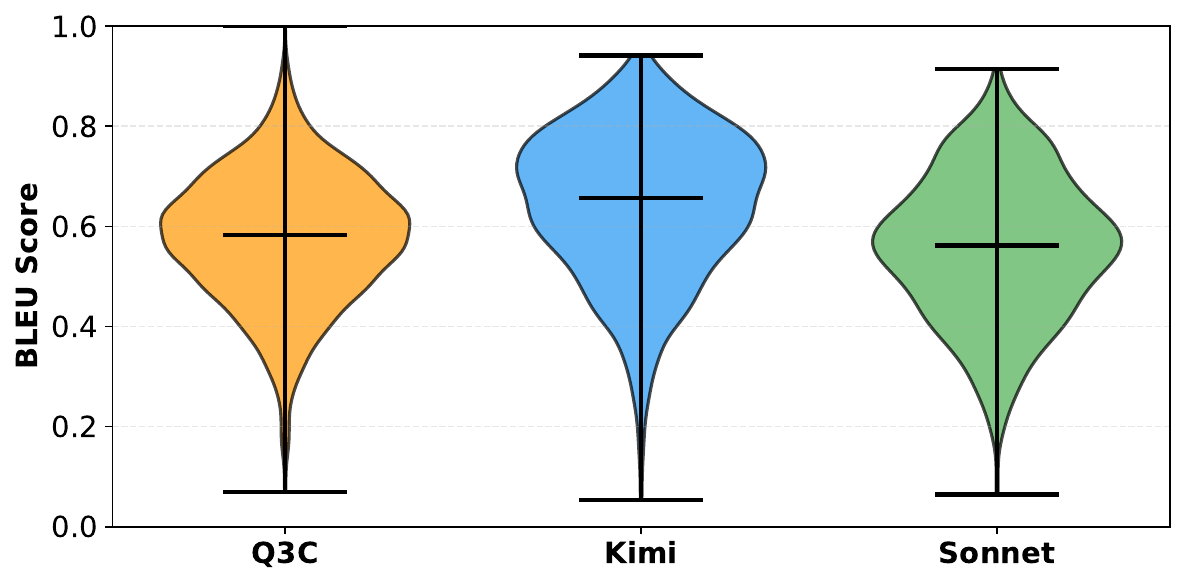}
    \caption{Scratchpad entries are highly similar to reasoning content that
already appears in \texttt{BashOnly}. For each scratchpad entry, we
compute its BLEU score against the corresponding reasoning content from
the \texttt{BashOnly} setup. Across all three actors, the distribution
is concentrated at relatively high scores, indicating that the
scratchpad often externalizes reasoning that the model would have
produced anyway rather than introducing substantively new intermediate
state.}
    \label{fig:scratchpad-bleu}
\end{figure}

\begin{figure}[t]
    \centering
    \includegraphics[width=0.75\linewidth]{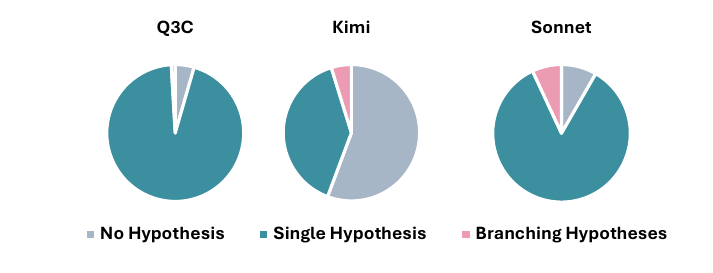}
    \caption{Hypothesis tracking rarely induces genuine branching behavior.
We categorize trajectories into three cases: no hypothesis-tracker use,
single-hypothesis use, and branching-hypothesis use. Across all three
actors, most trajectories either never use the tool or record only a
single hypothesis, while true branching over multiple alternative
hypotheses is rare.}
    \label{fig:hypotrack-pie}
\end{figure}

The two cognitive-scaffolding setups, \texttt{Scratchpad} and \texttt{HypoTrack}, have only
limited effects in our main evaluation. We interpret this result as specific
to the lightweight text-based scaffolds studied here. These tools provide an
explicit channel for recording intermediate reasoning, but they do not add
retrieval, memory management, new task information, or enforce a different
reasoning policy. Thus, they mainly test whether a simple text interface for
actor-written intermediate state changes behavior under otherwise similar
capabilities. In our setting, actors usually do not use these tools in ways
that substantially change their reasoning: under \texttt{Scratchpad}, actors often
produce content that closely matches reasoning already present in \texttt{BashOnly};
under \texttt{HypoTrack}, actors rarely maintain multiple competing hypotheses.
Taken together, these results suggest that lightweight text-based cognitive
scaffolding alone is insufficient to reliably alter actor reasoning behavior.

\noindent\textbf{Scratchpad mostly repeats reasoning that already
appears under \texttt{BashOnly}.}
To test whether \texttt{Scratchpad} introduces genuinely new reasoning,
we compare scratchpad content against the reasoning content produced in
the corresponding \texttt{BashOnly} trajectories. For each scratchpad
entry, we compute its BLEU score relative to the original
\texttt{BashOnly} reasoning content. Figure~\ref{fig:scratchpad-bleu}
shows that these scores are generally high across all three actors, with
many entries above 0.6. This indicates that scratchpad content is often
textually close to reasoning the model would already have produced
without the tool~\citep{lavie2010evaluating,chousa2025automatic,scarton2019estimating}. 
In other words, the scratchpad frequently acts as an
explicit channel for restating existing reasoning, rather than as a tool
that elicits substantially different intermediate thoughts or planning
structure.

\noindent\textbf{HypoTrack rarely induces multi-branch reasoning.}
The intended role of \texttt{HypoTrack} is to encourage the model to
manage alternative debugging hypotheses explicitly. To evaluate whether
this happens in practice, we classify trajectories into three cases:
\emph{no hypothesis}, where the tool is never called;
\emph{single hypothesis}, where only one hypothesis is recorded without
branching; and \emph{branching hypotheses}, where the model records
multiple alternative hypotheses for the same problem instance.
Figure~\ref{fig:hypotrack-pie} shows that true branching behavior is
rare. Most trajectories either do not use the tool at all or use it in
a single-threaded way by recording only one hypothesis. Thus, although
\texttt{HypoTrack} provides an interface for structured branch
management, actors seldom use it to explore multiple competing
explanations.

\section{How Actors Use the Provided Tools Across Architectures}
\label{appdx:tool-usage-patterns}

\begin{figure}[t]
    \centering
    \includegraphics[width=0.95\linewidth]{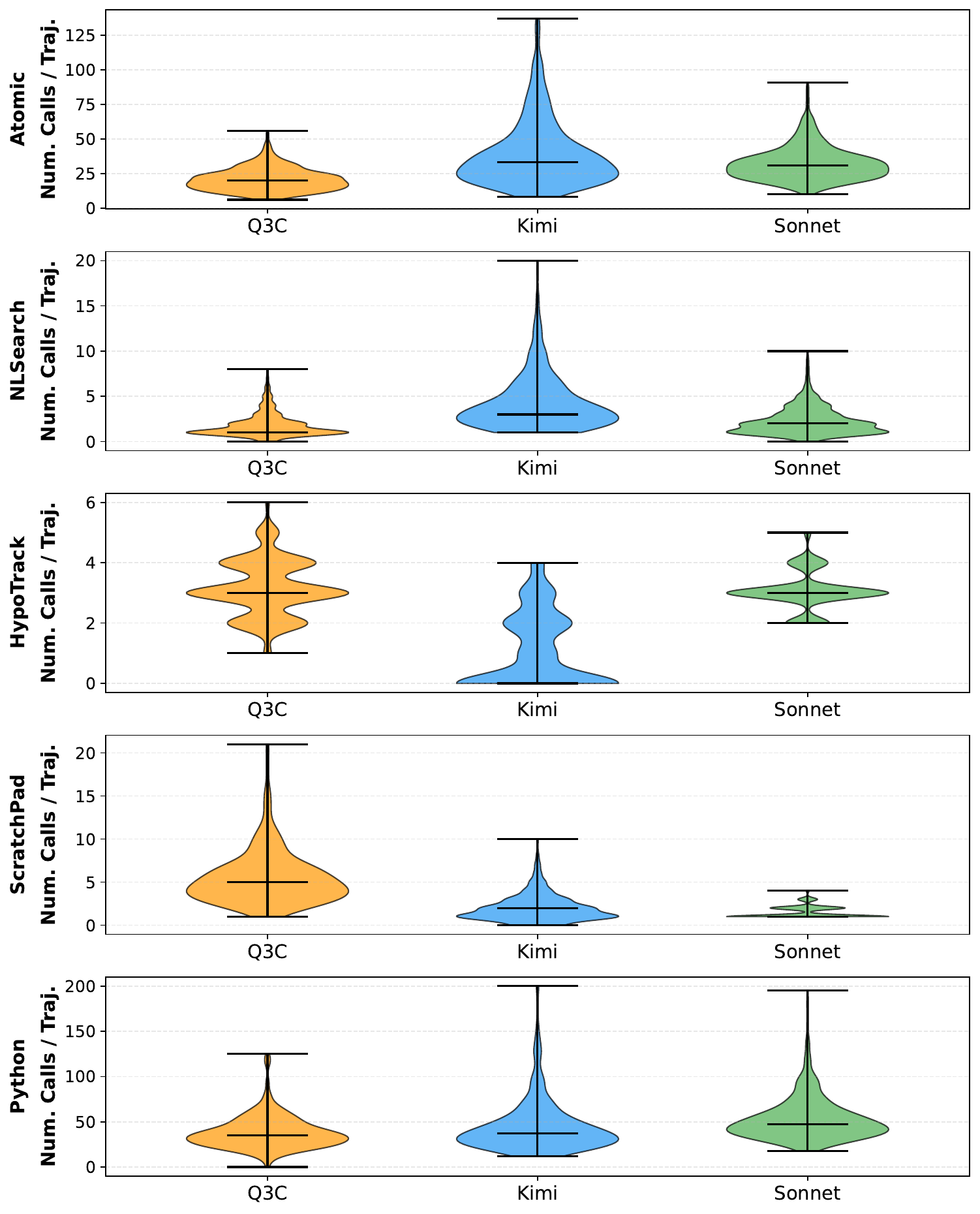}
    \caption{Distribution of tool-call frequency per trajectory under each
tool architecture. Each violin shows the number of calls to the
architecture-specific tool within a trajectory. Actors use
\texttt{Atomic} and \texttt{Python} frequently across trajectories,
whereas \texttt{NLSearch}, \texttt{HypoTrack}, and \texttt{Scratchpad}
are invoked much less often.}
    \label{fig:tool-call-freq}
\end{figure}

\begin{figure}[t]
    \centering
    \includegraphics[width=0.95\linewidth]{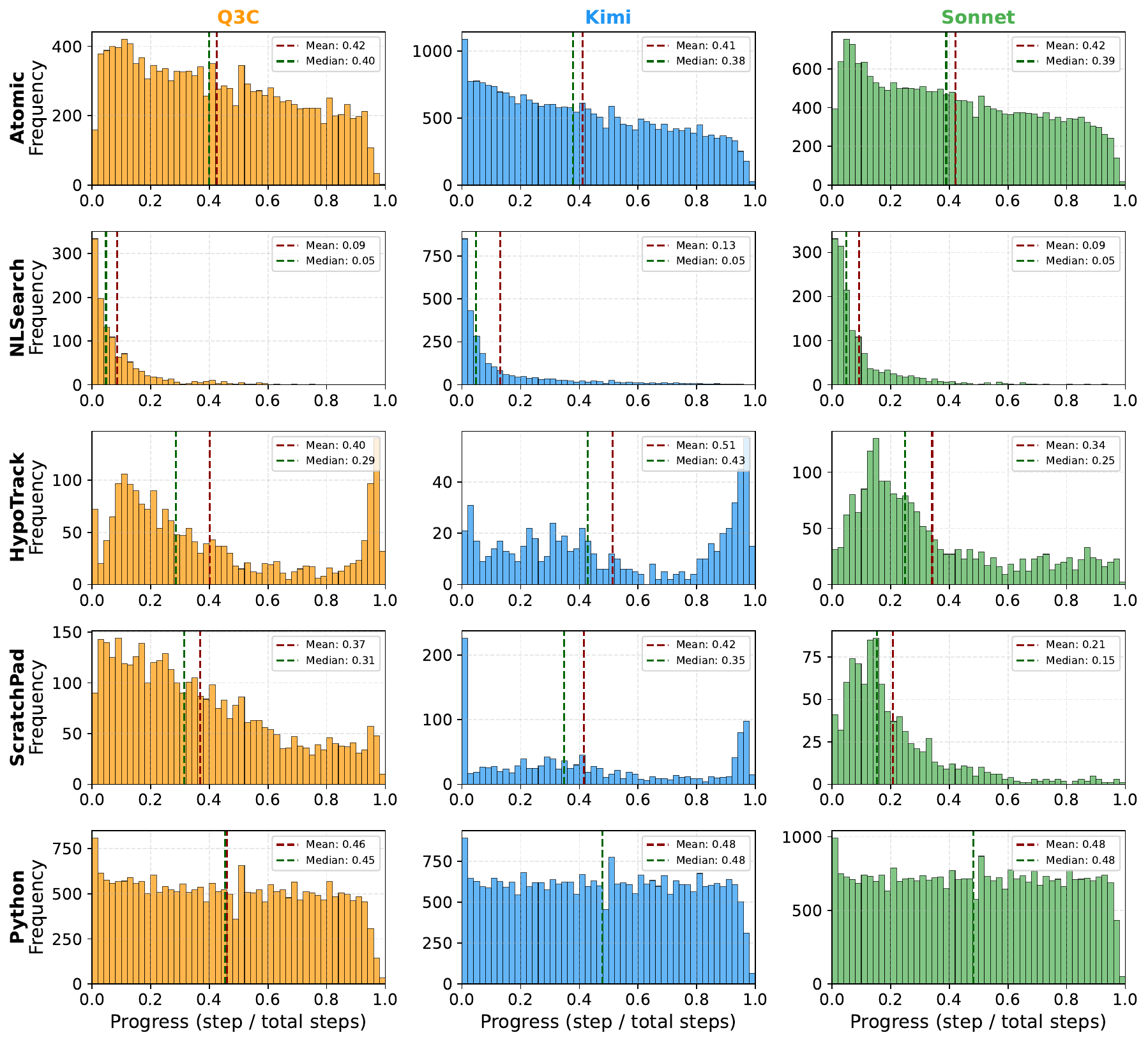}
    \caption{Distribution of tool-call locations within trajectories. For
each tool call, we compute its normalized location as
$\textit{step} / \textit{total trajectory length}$. \texttt{Atomic} and
\texttt{Python} calls are spread broadly across the trajectory,
\texttt{NLSearch} calls concentrate near the beginning, and
\texttt{HypoTrack} calls tend to appear near the beginning and the end,
consistent with initial hypothesis recording followed by a final status
update.}
    \label{fig:tool-call-loc}
\end{figure}

Tool usage differs substantially across architectures. Figure~
\ref{fig:tool-call-freq} shows the number of calls per trajectory for
each tool setup, while Figure~\ref{fig:tool-call-loc} shows when those
calls occur within a trajectory, measured as the step of a tool call
divided by the total trajectory length. Two patterns are clear. First,
actors adapt strongly to \texttt{Atomic} and \texttt{Python}: these
tools are used frequently across all three actors, indicating that they
become part of the main execution loop rather than serving as occasional
auxiliary tools. By contrast, \texttt{NLSearch} is used much less
frequently, which is consistent with its role as an exploration tool
rather than a general interaction interface. The cognitive-scaffolding
tools are also used relatively sparingly: \texttt{HypoTrack} is called
fewer than five times in most trajectories, and \texttt{Scratchpad} is
called fewer than ten times in most trajectories for
\texttt{Kimi-K2.5} and \texttt{Sonnet-4.5}.

Figure~\ref{fig:tool-call-loc} further shows that \texttt{Atomic} and
\texttt{Python} calls are distributed relatively evenly across the full
trajectory, consistent with their role in routine interaction. One
exception is that \texttt{Atomic} calls become slightly less frequent
late in the trajectory, likely because actors often switch back to
\texttt{bash} for test execution (which needs to execute the \texttt{python} bash command).
\texttt{NLSearch} concentrates near the beginning of trajectories, which matches its role
in repository exploration. \texttt{HypoTrack} calls cluster near
the beginning and the end: the beginning typically corresponds to
recording an initial hypothesis, while the end often corresponds to
updating the status of that same hypothesis. This pattern suggests that
actors seldom use \texttt{HypoTrack} to maintain multiple competing
lines of reasoning, which is consistent with the analysis in Appendix~
\ref{appdx:cog-scaffold-analysis}. Overall, these results suggest that
actors do adapt their tool use to the provided architecture, but the
degree and manner of adaptation differ substantially across tools.

\section{Detailed Discussion of Related Work}
\label{appdx:related}

\noindent \textbf{Empirical studies of coding agents.}
A growing body of recent work empirically studies coding agents, but
mostly from perspectives other than tool architecture. One line of
work is problem-focused, analyzing where coding agents succeed or fail
on tasks such as issue resolution, test generation, and patch
construction. Prior studies examine agent trajectories and error
patterns on software engineering benchmarks, analyze patch quality and
failure modes, and question whether strong benchmark results always
reflect genuine reasoning rather than memorization
~\citep{ceka2025understandingsoftwareengineeringagents,
majgaonkar2025understandingcodeagentbehaviour,
liang2025swebenchillusionstateoftheartllms}. A second line studies
how developers use coding agents in practice, including empirical
analyses of agent-assisted refactoring, developer adoption patterns,
and broader engineering practices around agent frameworks
~\citep{horikawa2025agenticrefactoringempiricalstudy,
watanabe2026useagenticcodingempirical,
wang2025empiricalstudyagentdeveloper}. A third line focuses more on
planning and failure behavior, for example by characterizing how
agent-based systems break down during automated issue solving
~\citep{liu2025empiricalstudyfailuresautomated}. Finally, recent work
also studies agent context and configuration, such as how developers
provide contextual instructions, configure coding agents, or maintain
agent-oriented repository files
~\citep{galster2026configuringagenticaicoding,
chatlatanagulchai2025agentreadmesempiricalstudy}.

Although these studies provide important empirical evidence about
coding-agent behavior, they do not consider tool architecture as a
design variable. As a result, they offer limited guidance on how the
organization and exposure of tools themselves shape agent behavior.

\noindent \textbf{Empirical studies of tool design.}
Another related line of work studies tool use more directly,
especially whether agents can identify and invoke the appropriate tool
in complex environments. Benchmarks such as
MCP-Bench~\citep{wang2025mcpbenchbenchmarkingtoolusingllm} and related
work evaluate tool-using agents under large toolsets, realistic APIs,
and multi-step tasks, emphasizing challenges in tool retrieval, schema
interpretation, and invocation accuracy
~\citep{fan2025mcptoolbenchlargescaleai}. However, this line of work
primarily treats tool design as a question of capability and correct
invocation. In contrast, our work asks whether, even when underlying
capabilities are similar, differences in interface architecture,
namely how capabilities are organized and exposed, can systematically
influence agent behavior.

\noindent \textbf{Agent non-functional properties.}
Recent work has also begun to examine non-functional properties of
agents, moving beyond aggregate task success. For example, prior work
studies efficiency, such as how turn-control strategies affect the
cost of coding agents~\citep{gao2025lessempiricalstudyturncontrol};
robustness, such as how tool-using agents behave under noisy or
perturbed environments
~\citep{wang2026agentnoisebenchbenchmarkingrobustnesstoolusing}; and
reliability, including broader calls for more systematic evaluation of
consistency and dependability in agentic systems
~\citep{rabanser2026towards}.
These works establish the importance of non-functional properties, but
they do not connect them to specific tool-architecture choices. Our
study addresses this gap by examining how tool architecture affects
efficiency, repeated-attempt consistency, and exploration.

\noindent \textbf{Tool design in coding agents.}
A substantial body of work on coding agents proposes concrete tool
designs, including repository navigation tools, structured search
interfaces, semantic retrieval systems, and agent frameworks built
around different interaction abstractions
~\citep{pan2026prometheuslonghorizoncodebasenavigation,yang2024swe,
liu2025reposcope,ouyang2024repograph,claudecode,
antoniades2024swe,openhands}. These systems are directly relevant to
our work because they motivate the representative tool architectures
considered in our study.

However, they typically introduce new tools as part of
end-to-end agent designs, often changing capability and interface at
the same time. As a result, although they show that particular tool
designs can improve task performance, they do not systematically
isolate the effect of tool architecture under similar capabilities.
Our work instead treats tool architecture itself as an experimental
variable and studies its effect on efficiency, consistency, and
exploration.

\end{document}